\documentstyle[11pt,aaspp4,flushrt]{article}

\def\ref{\par\noindent\hangindent=1in\hangafter=1}
\def\arcsec{{\tt ''}}			
\def\etal{et al.}			
\def\H0{$H_0$~= 75 \kms\ Mpc$^{-1}$}
\def\Halpha{H$\alpha$}

\def\kms{km s$^{-1}$}

\def\Mv{$M_V$}
\def\mv{$m_V$}

\def\reff{$R_{\rm eff}$}

\def\ub{$U\!-\!B$}

\def\bv{$B\!-\!V$}

\def\vi{$V\!-\!I$}

\def\ref{\par\noindent\hangindent 30pt}

\def\farcs{\hbox{$.\!\!^{\prime\prime}$}}

\def\v16{$\Delta V_{1-6}$}

\newcommand{\siiv}{Si~IV $\lambda$1400}
\newcommand{\civ}{C~IV $\lambda$1550}
\newcommand{\cii}{C~II $\lambda$1335}
\newcommand{\siii}{Si~II $\lambda$1526}

\newcommand{\Ms}{M$_{\odot}$}

\newcommand{\ergs}{erg~s$^{-1}$~cm$^{-2}$~\AA$^{-1}$}

\begin{document}

\clearpage

\thispagestyle{empty}	

\begin{center}

\begin{Large}

{\bf THE LUMINOSITY FUNCTION OF YOUNG STAR CLUSTERS 
IN ``THE ANTENNAE'' GALAXIES (NGC~4038/4039)$^1$} \\

\end{Large}

\end{center}

\begin{large}			

\bigskip
\bigskip

\centerline{BRADLEY C. WHITMORE, QING ZHANG$^2$, CLAUS LEITHERER, S. MICHAEL FALL, }

\centerline{Space Telescope Science Institute, 3700 San Martin Drive,
Baltimore, MD 21218}
\centerline{Electronic mail: whitmore@stsci.edu}

\bigskip
\bigskip

\centerline{FRAN\c COIS SCHWEIZER,}

\centerline{Carnegie Institution of Washington, Department of Terrestrial
Magnetism, 5241 Broad Branch Road, NW,}
\centerline{Washington, DC 20015}
\centerline{Electronic mail: schweizer@dtm.ciw.edu}

\bigskip
\bigskip

\centerline{and BRYAN W. MILLER}

\centerline{Sterrewacht Leiden, Postbus 9513, 2300 RA Leiden, The Netherlands}
\centerline{Electronic mail: bmiller@strw.LeidenUniv.nl}

\bigskip
\bigskip

Received:

\bigskip
\bigskip

\noindent $^{1}$ Based on observations with the NASA/ESA {\it Hubble Space 
Telescope}, obtained at the Space Telescope Science Institute, which is
operated by the Association of Universities for Research in Astronomy,
Inc., under NASA contract NAS5-26555.

\noindent $^{2}$ Also Johns Hopkins University, Charles and 34th St, Bloomberg Center, Baltimore, MD, 21218.

\clearpage

\centerline{\bf ABSTRACT}

The Wide Field and Planetary Camera 2 of the {\it Hubble Space
Telescope} has been used to obtain high-resolution images of NGC
4038/4039 that go roughly 3 magnitudes deeper in $V$ than previous
observations made during Cycle 2. These new images allow us to measure
the luminosity functions (LF) of clusters and stars over a range of 8
magnitudes ($-14 < M_V < -6$).  To first order the LF is a power law,
with exponent $\alpha = -2.12 \pm 0.04$.  However, using a variety of
different techniques to decouple the cluster and stellar LFs, which
overlap in the range $-9 \lesssim M_V \lesssim -6$, we find an
apparent bend in the young cluster LF at approximately $M_V = -10.4$.
Brightward of this magnitude the LF has a power law exponent $\alpha =
-2.6 \pm 0.2$ while faintward the slope is $\alpha = -1.7 \pm 0.2$.
The bend corresponds to a mass $\approx 1 \times 10^{5}$
M$_{\odot}$, only slightly lower than the characteristic mass of globular
clusters in the Milky Way ($\approx$ $2 \times 10^{5}$ M$_{\odot}$).
It is currently not feasible to determine the cluster LF fainter than
$M_V \approx -8$, where individual stars are likely to dominate.  The
stellar LF in the range $-9 < M_V < -6$ is much steeper, with $\alpha
= -2.9 \pm 0.1$, and is dominated by young red and blue supergiants.
The star clusters of the Antennae appear slightly resolved, with
median effective radii of $4 \pm 1$ pc, similar to or perhaps slightly
larger than those of globular clusters in our Galaxy. However, the
radial extents of some of the very young clusters (ages $<$ 10 Myr)
are much larger than those of old globular clusters (e.g., the outer
radius of Knot S exceeds 450 pc).  This may indicate that the tidal
forces from the galaxies have not had time to remove some of the outer
stars from the young clusters. A combination of the $UBVI$ colors,
\Halpha\ morphology, and GHRS spectra enables us to age-date the
clusters in different regions of The Antennae.  Star clusters around
the edge of the dust overlap region appear to be the youngest, with
ages $\lesssim 5$~Myr, while clusters in the Western Loop appear to be
5 -- 10 Myr old. Many star clusters in the northeastern star-formation
region appear to be $\sim$100 Myr old, with a LF in $V$ that has
shifted faintward by $\sim$1.0 mag relative to the younger (0 -- 20
Myr) clusters that dominate over most of the rest of the galaxy. A
third cluster population consists of intermediate-age clusters
($\sim$500 Myr) that probably formed during the initial encounter
responsible for ejecting the tails. A handful of old globular clusters
from the progenitor galaxies are also identified.  Most of these lie
around NGC 4039, where the lower background facilitates their
detection.  Age estimates derived from GHRS spectroscopy yield $3 \pm
1$ Myr for Knot~K (just south of the nucleus of NGC 4038) and $7 \pm
1$ Myr for Knot~S in the Western Loop, in good agreement with ages
derived from the $UBVI$ colors.  Effective gas-outflow velocities from
Knots S and K are estimated to be about 25 -- 30 \kms, based on the
above cluster ages and the sizes of the surrounding \Halpha\ bubbles.
However, the measured widths of the interstellar absorption lines
suggest dispersion velocities of $\sim$400 \kms\ along the lines of
sight to Knots S and K.

\bigskip


\clearpage

\centerline{\bf 1. INTRODUCTION}

One of the important early results from HST was the discovery of
young, compact, extremely bright star clusters in merging galaxies
(e.g., Holtzman \etal\ 1992, Whitmore \etal\ 1993).  A large number of
subsequent papers described similar objects in other mergers,
starburst galaxies, and even barred galaxies (e.g., Holtzman \etal\
1996, Stiavelli \etal\ 1998, Meurer \etal\ 1992, Conti \& Vacca 1994,
Barth \etal\ 1995; see Whitmore 1998 for a more complete summary).
The nearest example of a full scale, ongoing merger is NGC 4038/4039,
``The Antennae'' galaxies.  This represents perhaps our best chance
for understanding what triggers bursts of star and cluster formation
in mergers.  In Whitmore \& Schweizer (1995, hereafter Paper I), we
found that the luminosity function (hereafter LF) $\phi(L)$ for
point-like objects in NGC 4038/4039 was a power law, $\phi(L)dL
\propto L^{\alpha} dL$, with an exponent of $\alpha = -1.78 \pm 0.05$
down to a limiting magnitude of about $M_V = -9.3$. This power-law
shape agrees with the LF of Magellanic Cloud clusters and Galactic
open clusters (Elson \& Fall 1985) and the mass function of Giant
Molecular Clouds (Harris \& Pudritz 1994, Elmegreen \& Efremov 1997),
but differs from the LF of old globular clusters which is typically
lognormal, i.e., a Gaussian distribution with a FWHM of $\sim$3 mag in
a number vs. magnitude plot.  One of the possible explanations for
this apparent difference is that we did not observe faint enough to
see the turnover of the LF in NGC 4038/4039.  Our new WFPC2
observations of this system go roughly 3 mag deeper in $V$ than the
earlier observations, allowing us to check whether the power-law shape
of the LF continues to fainter magnitudes.

In Paper I we estimated that the mean effective radius (containing
half the light) of the young clusters in NGC 4038/4039 is 18 pc (for
$H_0$ = 50 km s$^{-1}$ Mpc$^{-1}$), although we also cautioned that
``the measurement of cluster sizes is severely compromised by the poor
image quality of the unrepaired telescope.'' As van den Bergh (1995)
pointed out, this mean effective radius is significantly larger than
that of normal globular clusters in the Milky Way, which average
around 3 pc.  Indeed, Meurer \etal\ (1995) measured typical effective
radii of 2 - 3 pc for young compact clusters in
nearby starburst galaxies with cluster populations similar to NGC
4038/4039; these authors suggested that the larger distance and more
severe crowding of clusters in NGC 4038/4039 had caused us to
overestimate the radii.  The repaired optics, availability of images
from the better sampled Planetary Camera instead of the Wide Field
Camera, and use of subpixel dithering all make the current data set
vastly superior to the earlier WF/PC-1 observations for the purpose of
measuring cluster sizes.

The current paper focuses on the LF, sizes, and ages of the clusters in
NGC 4038/4039, as determined from broad-band images in $UBVI$ and from
UV spectroscopy of two knots using the Goddard High-Resolution Spectrograph
(GHRS).  Future papers will discuss the ages and the mass function 
of the clusters in more detail (Zhang \& Fall, 1999), and will examine correlations with other
physical quantities such as neutral hydrogen, the radio continuum,
IR properties, X-ray emission, and the velocity field (Zhang \etal\ 1999).

In the following we adopt a Hubble Constant of
$H_0 = 75$ km s$^{-1}$ Mpc$^{-1}$,
which places NGC 4038/4039 at a distance of 19.2 Mpc, corresponding to
a distance modulus of 31.41 mag.  Note that this distance differs from
that in Paper I, where we adopted $H_0$ = 50 km s$^{-1}$ Mpc$^{-1}$.  At
the distance of 19.2 Mpc the projected scale is 1$''$ = 93 pc, and 1 pixel
on the Planetary Camera covers 4.23 pc, while one pixel on the Wide
Field Camera covers 9.26 pc.

\bigskip

\centerline{\bf 2. OBSERVATIONS AND REDUCTIONS}

\centerline{\it 2.1 WFPC2 Observations}

As a followup to the Cycle 2, pre-refurbishment observations of NGC
4038/4039 with the WF/PC-1, an extensive set of much deeper
observations were obtained with WFPC2 during Cycle 5 (proposal ID =
5962).  On January 20, 1996, four separate exposures were taken
through each of four broad-band filters, with total integration times
of 4500 sec in F336W (referred to as the ``U'' filter), 4000 sec in
F439W (``B''), 4400 sec in F555W (``V''), and 2000 sec in F814W
(``I'').  Photometric measurements obtained from these images were
converted to Johnson $U,B,V$, and Cousins $I$ passbands (note that
this, too, is a change from Paper I where the F785LP measurement was
converted to the Johnson $I$ passband). The foreground reddening in
the Milky Way toward NGC 4038/4039 is only 0.03 mag (Burstein \&
Heiles 1984) and has been ignored in this paper since it is negligible
compared to the uncertainty in the value of the internal reddening.  The
internal reddening itself is highly variable throughout the galaxy.
Two approaches are used below to minimize the effects of reddening due
to dust (\S 4.1 and \S 4.2). The first approach is to divide the
galaxies into regions with different levels of star formation and
dust, and work primarily on the regions with minor levels of reddening
and extinction.  The second approach is to use reddening-free Q
parameters to estimate and attempt to correct for extinction by dust.

The exposures in each filter were subpixel dithered to improve spatial
resolution and to flatten the background, using pairs of exposures that
were designed to be offset by 0.25$''$ (i.e. 2.5 pixels on the Wide
Field Camera [hereafter WFC] and 5.5 pixels on the Planetary Camera
[hereafter PC]) from each other. The actual offsets on the PC measured
using the STSDAS task {\it crosscor} varied from 5.21 pixels
in Y to 6.26 pixels in X, in both cases on the F336W image. The mean
offset for all the filters was 5.62 pixels. We note that the F336W
observations had the longest separation in time (6 orbits), which may
explain why they were the most discrepant from the
target value of 5.5 pixels. 

In addition to the long exposures, pairs of 60 sec subpixel dithered
exposures were taken through the F439W, F555W, and F814W filters in
order to recover the saturated peaks of the brightest regions in the
long-exposure images.  On January 13, 1996, narrow-band \Halpha\
images were obtained through the F658N filter designed to pass the
[\ion{N}{2}] $\lambda$6584 line at rest.  At the redshift of
NGC 4038/4039, this filter passes the \Halpha\ line, as desired. 
We used the same strategy as for the broad-band images, but
with a total exposure time of 3800~sec.  In addition, two subpixel
dithered images of Star-6 (see Paper I) of 30~sec duration each and centered
on the PC were taken to provide a good point-spread-function (PSF).
Additional images were taken with the Faint Object Camera (FOC), but are
not used in the present paper.  

Cosmic-ray tracks were removed using the STSDAS task {\it gcombine}
on each matched pair of images, and then offset pairs were combined
using the DRIZZLE software written by Fruchter \& Hook (1998). A pixfrac
parameter of 0.8 and scale parameter of 0.5 were used for the drizzling,
thus converting each $800\times 800$ image into a $1600\times 1600$ image.
Figure 1 shows an example of the separate and DRIZZLE-combined \Halpha\
images around Knot~G, along with a deconvolved image computed via the 
routine {\it acoadd} and a TINY TIM PSF (Krist 1995).  The advantage of
subpixel dithering is readily apparent:  the {\it acoadd} image clearly
shows the most structure.  However, since {\it acoadd} uses a non-linear
algorithm that tends to amplify noise, we have not used this image
for any measurements reported in the present paper.

After the subimages were drizzled together, bad pixels were removed
using the hot-pixel lists obtained from the STScI WFPC2 web site
and the task {\it warmpix}, and the saturated peaks of the brightest
stars and clusters were replaced with properly scaled portions of
the shorter-exposure images.

Photometric zeropoints were adopted from Table 28.1 of Version 3 of
the {\it HST Data Handbook} (Voit 1997), and color transformations from
Holtzman \etal\ (1995). Aperture corrections were non-trivial, since
the clusters are clearly resolved and vary in size. An average cluster
profile was determined from a number of isolated, high S/N clusters on
each chip, and this profile was used to determine the aperture
correction.  While this should be roughly correct for a typical,
relatively bright cluster, it should be recognized that the total
magnitude for any particular cluster may be off by a couple tenths of
a magnitude. An extreme case (Knot S) is discussed in \S 4.4.2.  This
magnitude uncertainty should have a very small effect on the LF
presented in this paper, and has essentially no effect on the colors,
since magnitudes within the same fixed apertures are used to form
color indices.  Aperture photometry was performed on all pointlike
objects.  The radii of the object aperture, the inner boundary of
the background annulus, and the outer boundary were 4/10/15 pixels
for the PC and 3/7/11 pixels for the WFC.  Note that these pixel triplets
correspond to values on the original WFPC2 images that are half these
numbers, since a scale of 0.5 was used in the DRIZZLE routine. Typical
aperture corrections are 0.90 mag on the PC and 0.60 mag on the WFC.
The fact that these corrections are
larger than the corresponding aperture corrections for stars by 0.2
mag (WFC) to 0.4 mag (PC) shows that the clusters are fairly well
resolved.  A comparison with the measurements of $V$ in
Paper I for 13 bright clusters shows a mean difference of 0.19 mag,
with the Paper I values being brighter. This reflects the
difficulty of obtaining accurate photometry from spherically
aberrated observations in a crowded field.  The dispersion of the
differences in $V$ is about 0.12 mag. The comparison for \vi\ color is
better with a mean difference of only 0.01 mag and a dispersion of 0.06 mag.

The task {\it daofind} from the DAOPHOT package (Stetson 1987) was
used to identify point-like objects (stars and clusters) from a
median-divided DRIZZLE-combined $V$ image  (see Miller et al.\ 1997 for a
discussion of the advantages of using a median-divided image). A comparison
using aperture photometry from the two dithered images provided one
last screening against the remaining cosmic rays, hot pixels, and chip
defects.  Completeness tests were performed for seven levels of the
background on both the PC and WFC by using the task {\it addstar} in
DAOPHOT to add artificial objects derived by using isolated, high S/N
clusters from the same image. Figure 2 shows the resulting completeness
fractions. The completeness thresholds used in this paper are
defined as the magnitude where only half the artificial objects
were identified.  Corrections for non-optimal charge transfer efficiency
on the CCDs of WFPC2 were made using the formulae by Whitmore \& Heyer
(1997). 

The main advantages of the new observations with WFPC2 are the much
improved spatial resolution and the correspondingly fainter thresholds
for object detection, resulting in the identification of $\sim$14,000
point-like sources as compared to $\sim$700 in Paper I.  The 20-fold
increase in detected objects is partly due to the fact that individual
stars become an important component faintward of $M_V \sim -9$, just beyond
the absolute-magnitude threshold reached by our previous WF/PC-1
observations. Hence, in the outskirts of some of the richest clusters
we are now able to study the clusters on a
star-by-star basis, at least for the top few stellar magnitudes.
Unfortunately, this also means that the problem of distinguishing between
stars and clusters becomes a major issue, which will be discussed at
length in the present paper.

Tables 1 through 4 gives the relative position, chip number, absolute magnitude
\Mv\ (assuming a distance modulus of 31.41), color indices,
concentration index \v16\ (i.e., the difference in $V$ magnitudes
measured in a 1 pixel and 6 pixel radius aperture, which provides a
rough determination of the cluster size), and previous identification
number (if it exists) from Paper I for the brightest 50 young cluster
candidates, 25 intermediate-age clusters, 11 old globular clusters,
and 13 foreground stars.

\bigskip

\centerline{\it 2.2 GHRS Ultraviolet Spectroscopy}

Ultraviolet spectra of Knots~S and K were obtained with the 
GHRS using Grating G140L on 1996 May 24. (See Rubin \etal\ 1970
for the original letter designations of the various knots.)  Knot~S was
initially acquired with the Faint Object Spectrograph (FOS) using a
3-stage peak-up, followed by an automatic small-angle maneuver to move
the center of this giant cluster into the GHRS Large Science Aperture
($1\farcs7 \times 1\farcs7$). The centering accuracy is about
$0\farcs35$, which corresponds to 1.5 GHRS science diodes. This in turn
translates into a wavelength uncertainty of $\sim$1~\AA. The science
exposure was centered at a wavelength of 1460~\AA\ and covered 286~\AA\
in wavelength.  The exposure time was 45~min.
Immediately following the exposure for Knot~S, a blind offset
was performed to Knot~K, whose separation from S was measured on
our WFPC2 images. A GHRS spectrum of K was taken with the same
instrument configuration as for S, but with an exposure time of 50~min.

Knot~S is dominated by a giant star cluster denoted as \#405
in Paper~I, and is listed as the second-brightest young cluster in
NGC4038/4039 in Table 1.  This cluster has a radius in excess of
450 pc (\S 4.4.2). The nearest major star cluster is \#430, which is
the 6th brightest young cluster in Table 1. It is 1.2 mag fainter in
$U$ and located $2\farcs3$ to the northeast of \#405, whence it should
not provide any measurable contamination within the
$1\farcs7 \times 1\farcs7$ aperture.  On the other
hand, Knot~K consists of two bright young clusters (\#442 and \#450,
the 3rd and 4th brightest clusters in Table 1) that are separated by
only $0\farcs7$, whence the recorded spectrum contains light from both
clusters. The two clusters of Knot~K have nearly identical colors and
formed, therefore, probably at nearly the same time.

The two cluster spectra were analyzed with standard routines from the
IRAF/STSDAS package. We found it unnecessary to recalibrate the
{\it calhrs} pipeline products.  After combining wavelength and flux data,
we rebinned the wavelength scale to the restframe of NGC~4038/4039
($v_{\rm{hel}} = 1650$~\kms; Paper I). Then the spectra were smoothed
with a 5 pixel ($\sim$1.0~\AA) boxcar filter and normalized to unity
by dividing them by a low-order spline function that had been fit to
line-free regions in the continuum.  Wavelengths, equivalent widths,
and line widths were measured in the normalized spectra using the {\it
splot} package by fitting Gaussians to the observed profiles.
The analysis of the spectra is presented in \S 4.5.

\bigskip

\centerline{\bf 3. GENERAL APPEARANCE}

Most of the major morphological features of The Antennae were
discussed in Paper I, based on WF/PC-1 observations.  Figure 3 shows a
``true-color'' image of The Antennae where the $U+B$ image is
displayed in blue, the $V$ image in green, and the $I$ image in red.
Figure 4 shows the same image as Figure 3 except that the \Halpha\
image replaces the $I$ image for the red display.  Figure 5a shows a
$V$ image of The Antennae with the apparent locations of various
regions of interest marked.  The lettering follows the original
definitions of most of these regions by Rubin \etal\ (1970) and in
Paper I.  Figure 5b identifies the brightest young cluster candidates,
intermediate-age clusters, old globular clusters, and foreground
stars.  Enlargements of particularly interesting regions are included
in Figure 6 ($UBVI$ image) and Figure 7 (\Halpha\ image).  Note the
resemblance of the loops near the center of NGC 4039 (Figure 6, region
A) to the loops seen around NGC 7252 (Schweizer 1982), on a scale
roughly 20 times smaller than in NGC 7252. We may be seeing stars and
gas being funneled into the center of the galaxy.

One of the most impressive features of Figure 3 is the fine structure
of the dust filaments. While many of the filaments are found near
regions of active star and cluster formation, there are also filaments
in regions with little or no star formation.  In particular, note the
long, completely dark filament in the southeast quadrant of Figure 3.
Conversely, there are many regions of active star formation with
little dust, based both on their appearance and the fact that many of
these clusters have relatively unreddened colors.
It appears that while regions of dense
gas and dust are required to fuel star formation, once this begins the
dust is fairly rapidly blown out or removed in some other way, leaving
the regions relatively free of dust.  We note that the most active
regions of star formation appear to lie on the edges of the overlap
region (e.g., regions B, C, D, E and F in Fig.~5). We also note that
the northeastern star formation region appears to extend behind the
overlap region, with its edge being visible along a line from region F
to region B (see Figs.~3 and 5a).  The northern part of NGC 4038 is
relatively symmetric and does not appear to differ dramatically from a
normal spiral galaxy, although it does feature a wider disk toward the
northeast where the clusters appear to be more spread out. This
suggests that this galaxy is at a relatively early stage of merging
and violent relaxation has yet to occur.

Perhaps the most dramatic thing about the \Halpha\ image of Figure 4
is the fact that one can immediately age-date the star formation in
the various regions, since \Halpha\ emission requires the presence of
O and B stars to ionize the gas and these stars last less than
$\sim$10 Myr. The regions immediately adjacent to the overlap region
appear to be the youngest, since the \Halpha\ bubbles are quite small
(see regions C and D in Fig.~7).  The clusters in the Western Loop
appear to be slightly older, since the \Halpha\ bubbles are larger
(see regions R+S+T in Fig.~7).  Finally, the northeastern star
formation region (around regions N and P) and the inner tail of NGC
4039 (region AA) are somewhat older still since they display little or
no \Halpha\ emission (best seen on Fig.~4).  Another interesting
result is that only large cluster complexes appear capable of blowing
large \Halpha\ bubbles, the largest of which measure more than 1 kpc
in diameter (see region G in Fig.~7).  Isolated stars and small clusters have
\Halpha\ and continuum images that are roughly coincident.

Many of the cluster complexes, such as region T (Fig.~6), contain dozens
of clusters.  However, the second brightest cluster of The Antennae,
Knot~S (Fig.~6), appears to be an exception, since it is the dominant
cluster of its region.  Note the extensive halo of stars
(typically with $-7 < M_V < -9$) surrounding this cluster
out to nearly 500 pc, more than twice the tidal radius of the largest
globular clusters in the Milky Way and M31 (\S 4.4). These extensive
outer envelopes of stars are seen only around the brightest clusters
and cluster complexes, presumably because these are also the youngest
stellar aggregates ($\sim$10 Myr).

\bigskip

\centerline{\bf 4. ANALYSIS}

\centerline{\it 4.1 Separating Stars and Clusters}

As briefly discussed in \S 1, Whitmore \& Schweizer (1995) found that
the LF of clusters in NGC 4038/4039 was a power law, $\phi(L)dL \propto
L^{\alpha} dL$, with $\alpha = -1.78 \pm 0.05$ down to a limiting absolute
magnitude of about $M_V = -9.3$. This is unlike the LF of old globular
clusters which  has a Gaussian profile with a peak around $M_V =
-7.2$ and a width of $\sim$1.4 mag.  Several possible explanations
for the apparent difference were discussed in Paper I, one being that
we had not observed faint enough to see the expected turnover.  The
new WFPC2 observations of NGC 4038/4039 go roughly 3 magnitudes deeper
than the earlier observations, allowing us to check whether the
apparent power-law shape of the cluster LF continues to fainter magnitudes.
Unfortunately, individual stars can have absolute magnitudes as bright as
$M_V \approx -9$, making the determination of the cluster LF at fainter
magnitudes more difficult.

While it is possible in principle to separate most stars and clusters
based on their apparent sizes, the crowded nature of the imaged field
and the fact that the clusters are only slightly resolved results in
some misidentifications.  In practice, a more reliable method of
identifying definite star clusters is to use a cutoff of $M_V < -9$
mag, which is brighter than all but a few of the brightest stars
(Humphreys 1983). The top panels in Figure 8 show the concentration
index \v16, the \vi\ vs. \Mv\ color-magnitude diagram, and the \ub\
vs. \vi\ color-color diagram for the cluster-rich regions on the PC
(i.e., regions T, S, R, and \# 13; see Paper~I for the latter
identification).  The evolutionary path of a Bruzual-Charlot (1996)
cluster model with solar (Z = Z$_{\odot}$) metallicity and ages marked
in logarithmic years is shown for reference.  The choice of
Z$_{\odot}$ is mainly for illustrative purposes, although this value
is typical of the few cluster metallicity estimates that have been
made in similar systems (e.g., Schweizer \& Seitzer 1998).

We find that the definite clusters (i.e., objects brighter than
$M_V = -9$ mag) show a peak in their concentration-index histogram
at \v16~$\approx 3.0$, colors that fall in a fairly
small range of $0.0 \lesssim V-I \lesssim 0.6$, and locations in the
color-color diagram that define a tight clump with a mean age of about
10 Myr, based on the Bruzual-Charlot cluster models for $Z_{\odot}$. Only
about 10 of the objects (or $\sim$5\%) fall outside the tight clump and are
probably individual stars (e.g., the mean value of 
$\overline{\Delta V}_{1-6} $= $2.20 \pm 0.14$,
with a scatter of 0.42
for these 10 objects, while $\overline{\Delta V}_{1-6}$ = $2.91 \pm 0.04$,
with a scatter of 0.48 for the 163 objects falling near the Bruzual-Charlot
models).
However, since the primary goal of the
present paper is to determine the LF to as faint a magnitude as
possible, we need to push beyond the $M_V = -9$ limit if possible.

The middle and bottom sets of panels in Figure 8 include all objects
brighter than  $M_V = -6$  for the cluster-rich and cluster-poor (i.e., west
of R, S, and T) regions, respectively, on the PC.  A comparison of the
color-magnitude diagrams in the two regions shows that while the
cluster-rich regions have a large number of objects brighter than $M_V
= -9$, the cluster-poor regions have only two, and hence are likely
to be dominated by individual young stars. In addition, most of the
objects in the cluster-rich regions fall in a tight clump in the
color-color plot, while the objects in the cluster-poor regions are
more spread out and have relatively few members in the clump defined
by the clusters in the top panel.  Finally, and perhaps most
importantly, the distributions of concentration indices are
dramatically different (i.e., there are almost no objects with
\v16~$\approx 3.0$ in the cluster-poor regions, unlike the top panel
where the peak at 3.0 indicates that the clusters are partially resolved),
although there is considerable overlap in these histograms.
Hence, in principle it should be possible to separate the stars from
the clusters in a statistical sense by using the concentration-index
and/or color information.

\bigskip

\centerline{\it 4.2 The Cluster LF}

{\it 4.2.1 The Total LF of Point-Like Objects in NGC 4038/4039}

Figure 9 shows the raw and completeness-corrected (but not reddening
corrected, see \S 4.2.2) LF for the objects on all four chips, as well as
on the PC alone.  The 50\% completeness limits are indicated by
arrows.  The top panel shows the Number vs. \mv\ plot (in which a
population of old globular clusters would have roughly a Gaussian
distribution), the middle panel shows the log(Number) vs. \mv\ plot
(where a power law is a straight line), and the bottom panel shows the
log(Number) vs. \mv\ plot for the PC objects alone.  The logarithmic LF
for all the point-like objects on all four chips is reasonably linear, with
a slope (power-law exponent) corresponding to $\alpha = -2.12 \pm 0.04$ when
fit to the completeness limit at $M_V = -7.0$.
This slope is steeper than the value found in Paper I ($\alpha = -1.78 \pm
0.05$), probably because there the spherical aberration made
it difficult to detect the fainter objects and no completeness
corrections were applied.  We note that the values of $\alpha$ on all
four chips are in relatively good agreement with each other ($\alpha = -2.17$,
$-$2.25, $-$2.03, and $-$2.12 for Chips 1 through 4, respectively).

The relatively good linear fit over more than  five magnitudes is somewhat
surprising, since we believe that the object counts at fainter magnitudes
are dominated by stars (see \S 4.1).  However, the bright end of the LF
from $M_V = -12.8$ to $-$10.4 is actually slightly steeper (i.e.,
$\alpha = -2.45 \pm 0.21$) than the overall LF, whereas the middle range
from $M_V = -10.4$ to $-$7.4 is slightly flatter (i.e.,
$\alpha = -2.09 \pm 0.02$).  While this slope difference is significant
only at the 1.7$\sigma$ level, we will present further evidence
below for the reality of this apparent bend in the LF at $M_V \approx -10.4$.

The power-law fits employed in the present paper are based on
least-squares fits weighted by the inverse of the variance
in each bin. We also performed a variety of simultaneous
two-slope power-law fits to quantify better the reality and position
of the apparent bend in the LF. The total LF for all objects shown
in Figure 9 is equally well fit by a single or a two-slope power law
(i.e., the reduced $\chi^2$ values  are essentially identical).

\bigskip

{\it 4.2.2 Decoupling the LFs of Clusters and Stars}

In an attempt to separate the cluster LF from the LF of the young stars
in NGC 4038/4039, we have used four different methods. In essence, these
methods consisted of: 

(1) restricting the sample to the part of the LF brighter than $M_V = -9$,
where clusters dominate (i.e., the approach used in \S 4.1),

(2) estimating the stellar LF from the cluster-poor regions, as shown in
the bottom panel of Figure 8, and subtracting this from the total LF to
determine the cluster LF,

(3) using the concentration (i.e., size) indices to attempt to isolate
the clusters, and

(4) estimating the stellar LF based on the color information and subtracting
it from the total LF to determine the cluster LF.

As we shall see, all four methods give reasonably similar results.

We begin by considering objects on the PC alone, since the better spatial
sampling of the PSF there makes it easier to separate the stars from 
barely resolved clusters. The bottom panel of Figure 9 shows the LF for
all objects on the PC.  Restricting our attention to the objects
brighter than $M_V = -9$ yields a value of $\alpha = -2.17 \pm 0.06$.
We again see a hint of an apparent bend at $M_V \approx -10.4$,
with  $\alpha = -2.95 \pm 0.29$ for the bright end of the LF and
$\alpha = -2.07 \pm 0.04$ for the faint end.

Figure 10 shows the completeness-corrected LF for the cluster-rich
regions on the PC, as well as the LF for the objects in cluster-poor
regions.  There are obvious differences between the two LFs, with a
steep LF and only a single object brighter than $M_V = -9$ in the
cluster-poor regions, and a much flatter distribution and 184 objects
brighter than $M_V = -9$ in the cluster-rich regions.  The power-law
exponent is $\alpha = -2.01 \pm 0.11$ when the LF is fit with a single
power law from $M_V = -11.4$ to $-$7.8 for the cluster-rich regions, and
$-2.92 \pm 0.09$ when the LF is fit from $M_V = -9.0$ to $-$6.0 for the
cluster-poor regions.  A two-slope power law  provides a much better fit
for the cluster-rich regions, with the parameters shown on the
figure. The slopes brightward and faintward of the bend now differ by
4$\sigma$.  The bend itself occurs at $M_V = -10.05 \pm 0.24$, brightward
of where we expect contamination by stars or completeness corrections to
become significant.

We can improve the statistics on a possible bend in the LF by using
the same approach on all four chips, as shown in Figure 11. Figure 11b
displays the LF for the cluster-rich regions only, while Figure 11a
displays the total LF for all regions of NGC 4038/4039 for comparison.
In Figure 11b, the dotted line shows the resulting cluster LF after the
LF from the cluster-poor regions is used to estimate the stellar LF and
subtract it from the total (normalizing the two by assuming that stars
dominate at $M_V = -7.4$).  The adopted normalization at $M_V = -7.4$
is not meant to imply that all objects at this absolute magnitude are
stars, but only to provide a lower bound to the possible cluster LF.
The effect of the subtraction on the cluster LF is relatively small out to
$M_V = -8.0$.  The resulting slope of the cluster LF faintward of the bend
is therefore bracketed between $\alpha = -1.76 \pm 0.03$ and $\alpha =
-1.59 \pm 0.06$.  The bend occurs at $M_V = -10.19 \pm 0.27$ in the former
case and at $M_V = -10.40 \pm 0.25$ in the latter case.  The change in
slope at $M_V \approx -10.4$ represents a $\sim$4$\sigma$ difference in
both cases.

Figure 12 shows the same diagram as Figure 11, but with fits using a
Schechter (1976) function to fit the observed LFs.  This function
provides similar or even slightly better fits to the data, with its
knee roughly 1 mag brighter than the bend found using the two-slope
power laws.

The third method for decoupling the stellar and cluster LFs uses
size information.  If we restrict the sample to objects on the PC with
\v16~$< 2.5$ (i.e., primarily stars), we find a steep stellar
LF with corrected $\alpha = -2.42 \pm 0.06$ for a fit in the range from
$-10.4 < M_V < -7.4$.  If we restrict the sample to objects on the PC
with \v16~$> 2.5$ (i.e., primarily clusters), as shown in Figure 11c, we
find a shallower cluster LF with $\alpha = -1.93 \pm 0.04$ in the same range.
The two-power-law fit ($\chi^2$ = 2.6) provides a much better fit
to the data than the single-power-law fit ($\chi^2$ = 3.8) does,
with the bend occurring at $M_V = -10.25 \pm 0.38$.

The fourth method for decoupling the stellar and cluster LFs uses 
color information  to isolate the stars.
Figure 8 shows that when we isolate the definite clusters
by including only the M$_V$ $<$ --9 objects, the resulting
color-color diagram shows a tight clumping at a location
in agreement with Bruzual-Charlot cluster models of $\approx$
10 Myr age. The bottom panel of Figure 8  shows that many individual
stars have very different colors.  We can use this fact to identify the
objects that  are almost certainly stars, and then subtract them from
the total to estimate the number of clusters.  This procedure only
provides an upper limit to the number of clusters, since some fraction of
the individual stars presumably have colors that lie in the cluster part
of the color-color diagram.

The Antennae contain a large amount of dust, especially in the
``Overlap Region'' between the two galaxies (see Fig.~5a). Extinction
from this dust affects the LF, while reddening affects
age estimates.  We can attempt to correct for the effects of dust by
using reddening-free ``Q'' parameters (Becker 1938; Johnson \& Morgan
1953) to estimate the intrinsic cluster colors, allowing us to estimate
the extinction. The technique is described briefly below.

We employ the galactic reddening law by Mathis (1990).  With the
$UBVI$ colors, three independent reddening-free $Q$ parameters can be
defined (cf.\ Mihalas \& Binney 1981) as
\begin{eqnarray}
   Q_1=(U-B)-0.72(B-V),\\
   Q_2=(B-V)-0.80(V-I),\\
   Q_3=(U-B)-0.58(V-I)
\end{eqnarray}

With these parameters, each object can be plotted on a $Q$--$Q$ plot
and compared with stellar population synthesis model tracks of cluster
evolution (e.g., Bruzual \& Charlot 1996) to determine its age and
intrinsic colors.  The difference between the observed and intrinsic color,
$E(B-V)$, is used to determine the extinction corrections for the
various passbands.

Due to the nature of the cluster-model tracks and the observed scatter
of data points, we have adopted the $Q_1-Q_3$ plot as the primary tool
in our attempt to decouple the stars from the clusters using color
information.  Figure 13a shows a two-color diagram for part of the
northeastern star formation region and the corresponding $Q_1-Q_3$
plot. Figure 13b shows the same plots for the cluster-rich regions on
the PC. The \ub\ vs.\ \bv\ diagrams show that the regions are
relatively dust free, since most of the objects fall near locations
only slightly displaced from the Bruzual-Charlot model tracks (solid
line).  A typical value of $E(B-V)$ is 0.3, with typical values for
A$_V$ = 1.0.  Small crosses with corresponding labels indicate the
values of log(age) along the Bruzual-Charlot models.  A comparison of
Figures 13a and Figure 13b indicates that there are a number of clusters
with ages $\approx$ 100 Myr (i.e., \ub~$\approx -0.3$, \bv~$\approx
0.1$) in the northeastern star formation region, but essentially no
clusters in this age range for the cluster-rich regions on the PC.
More about this in \S 4.3 . 

Besides the BC96 cluster-model tracks, loci for various types of stars
are also included on the diagrams.  As shown in the bottom panels of
Figures 13a and 13b, stars tend to be located to the upper right of
the evolutionary tracks for clusters, although there is some overlap
between cluster models and stars (especially blue supergiants).  Data
points that are not located in the regions of stars or clusters
(within the errors) are considered noise and are excluded from the
sample.

Figure 14 shows similar plots for objects in the Overlap Region (see
Fig.~5).  As expected, the reddening due to dust moves the data points
off the BC96 cluster-model track in the \ub\ vs. \bv\ diagram.  A
typical value of $E(B-V)$ is 0.8.  We find that most clusters in the
Overlap Region have ages $<$20 Myr, with no significant population of
clusters with ages $\sim$100 Myr, similar to the results from Figure
13b.  In addition, the offset from the model track in the $Q_1-Q_3$
plot for the overlap region suggests a problem in either the galactic
reddening law used to define the $Q$ parameters, or the stellar
evolution models.  However, a comparison between the Leitherer \&
Heckman (1995) models and the BC96 models results in almost identical
tracks up to approximately log{\it t} = 6.9.

Figures 13 and 14 show that many individual stars should be identifiable by
their colors, as was also apparent from Figure 8. Especially in the
$Q_1-Q_3$ plots, many types of stars lie well beyond the cluster-evolution
tracks.  Hence, we can estimate the number of stars by focussing on
objects with values of $Q_3 > 0.0$ and $Q_1 > 0.0$.  Using this
technique, we find that at least 40\% of the objects in NGC 4038/4039
are stars, which have subsequently been removed from Figures 13 and 14.
This estimated percentage is a lower limit since some stars will have
colors that make it impossible to distinguish them from clusters.  Subtracting
the resulting LF 
from the total LF we obtain an upper limit to the cluster LF.
Figure 11d shows the result after the stars identified from the $Q_1-Q_3$
plot have been subtracted.  Once again a two-slope power law provides a
better fit than a single power law, with a 5$\sigma$ difference in the slopes.

Figure 11 shows that all four methods for decoupling the stellar and
cluster LFs lead to cluster LFs with a bend at $M_V \approx -10.4$.
Even the total LF, to which stars are likely to contribute significantly
in the $M_V = -7$ to  $-$8 range, shows a hint of this bend.  Based on
these results, we conclude that the LF for young star clusters in
NGC 4038/4039 is best represented by two power-law segments with indices
of $\alpha = -2.6 \pm 0.2$ in the range $-12.9 < M_V < -10.4$ and
$\alpha = -1.7 \pm 0.2$ in the range $-10.4 < M_V < -8.0$, respectively.
It seems not possible at present to determine the cluster LF beyond
$M_V = -8.0$, where stars are likely to dominate.

Figure 15 shows the dereddened and completeness-corrected LF for the
cluster-rich regions on the PC. This compares fairly well with the top
panel of Figure 10, supporting our claim that the extinction by dust
is relatively small in most of the cluster-rich regions, and hence
does not affect the LF very much. The bright end of the LF is still
quite steep ($\alpha = -2.53 \pm 0.29$), the bend near $M_V \approx -10.4$
is more apparent, and the portion from $M_V = -10.4$ to $-$8.5 is flatter
($\alpha = -1.47 \pm 0.11$), presumably due to the removal of stars by
the $Q$--$Q$ method discussed above.  Unfortunately, with typical values of
$A_V \approx 2$ -- 3 mag for clusters in very dusty regions, it seems
nearly impossible to obtain reliable cluster LFs in these regions.

Could the slight bend at $M_V \approx -10.4$ be the precursor to the
observed peak at $M_V \approx -7.2$ in the LF of old globular clusters
in elliptical galaxies (e.g., Whitmore 1997)?  Based on our results in
\S 4.3, a typical age for the young clusters in our sample appears to
be about 10 Myr.  The Bruzual \& Charlot (1996) models would then
predict that the clusters should fade by about 5 mag in $V$ by the
time they are 15 Gyr old (Whitmore \etal\ 1997, Fig.~18), rather than
by the 3.2 mag difference between $M_V = -10.4$ and $-$7.2.  The
agreement is better if we add an extinction correction of 1 magnitude
to the bend (see \S 4.2.2), resulting in a 4.2 mag difference.

A rough estimate of the mass corresponding to the bend in the LF can
be made in a similar manner. Again adopting a typical age of 10 Myr, 1
mag of extinction, and using the solar metallicity BC96 models leads
to a mass of $1 \times 10^{5}$ M$_{\odot}$ at the bend of the LF. This
is somewhat lower than the characteristic mass of globular clusters in
the Milky Way ($ 2 \times 10^{5}$ M$_{\odot}$; corresponding to the
peak of the luminosity function for M/L$_V$ = 3).  Given various
uncertainties in these simplistic calculations, such as the exact
location of the bend, the age and metallicity to use for the
calculation, uncertainties in the distance, and uncertainties in the
Bruzual-Charlot models themselves, these estimates are probably
compatible with the values expected of Milky Way globular clusters.

With the benefit of hindsight we note that the cluster LFs in NGC 7252
(Figs.\ 20b and 20d in Miller \etal\ 1997) and NGC 1275 (Carlson
\etal\ 1998, Figs.\ 7 and 8) may also show a bend, but at about $M_V =
-9.5$ to $-$10.0.  In addition, Zepf \etal\ (1999) find tentative
evidence for a flattening in the cluser LF in the recent merger NGC
3256.  Though taken alone none of these cases of lower-luminosity
bends is compelling, in light of our new results for NGC 4038/4039 it
appears that they may be part of a trend. We also note that the bend
in the LF of the star clusters in NGC 4038/4039 is reminiscent of the
bend seen in the LF of H II regions observed in several nearby
galaxies (e.g., Kennicutt et al.\ 1989, Oey \& Clarke 1998).

Meurer (1995) has suggested that the shape of the LF may evolve from
an initial power law distribution to a log-normal distribution, due to
the spread in the ages of the clusters and the subsequent fading 
with time.  This important topic will be addressed in a 
paper by Zhang \& Fall (1999), who determine the cluster mass
function by removing the effects of fading and extinction. They
find that the luminosity function and mass functions are relatively
similar in shape.

{\it 4.2.3 Effects of Emission Lines on $UBVI$ Photometry}

Emission lines can affect the broad-band colors measured for very
young star clusters (e.g., Stiavelli \etal\ 1998).  For example, the
F555W filter passband contains strong emission lines at $\lambda\lambda$4861,
4959, and 5007 \AA, while the F814W filter passband does not contain any
strong lines (i.e., its short wavelength cutoff lies at
$\sim$7000 \AA, which excludes \Halpha).  The F439W filter (i.e.,
$B$) passband is relatively less affected since it covers the range
4000 -- 4700 \AA\ and misses the strong emission lines at
$\lambda\lambda$3727 and 4861 \AA.  Hence, it is mainly the $V$ magnitudes
of clusters in emission-line regions that are too bright, resulting in objects
being measured too blue in \vi\ and too red in \bv .

Figure 16 displays the \bv\ vs.\ \vi\ diagram for bright clusters on
the PC, with objects with strong \Halpha\ and weak (or absent)
\Halpha\ marked by different symbols.  The various lines show the
Bruzual-Charlot cluster-model tracks for three values of metallicity.
While the objects with weak emission lines tend to fall along the
Bruzual-Charlot tracks, those with strong emission lines are offset by
about $-$0.2 in \vi.  Hence, the presence of emission lines does
affect the measured broad-band colors that include $V$, but the effect
is relatively minor and its influence on the cluster LF should be
nearly negligible.  A comparison between LFs based on F439W, F555W,
and F814W observations shows that this is indeed the case.

\bigskip

\centerline{\it 4.3 Cluster Ages}

\centerline{\it 4.3.1 Evidence for Four Populations of Clusters in The Antennae}

Simulations of merging
galaxies suggest that star and cluster formation induced by the merger
should be spread over several hundred million years, rather than
happening in a rapid  burst (e.g., Mihos \etal\ 1993). Hence,
the resulting spread in cluster ages for ongoing mergers may provide
an opportunity to study the early evolution of star clusters in a
single system rather than having to intercompare several different
galaxies.

There is evidence for four different populations of star clusters in
The Antennae.  The youngest population is most efficiently identified
via the \Halpha\ images.  The presence of \Halpha\ emission alone
guarantees that the region contains clusters younger than $\sim$10
Myr, since O and B stars must be present to ionize the gas.  The size
of the \Halpha\ bubble can provide a further discriminant.  In the
cluster complexes bordering the Overlap Region of The Antennae (e.g.,
regions B, C, D, F; see Figure 5b), the \Halpha\ and continuum images
appear similar.  Apparently there has not been enough time for the
complexes to blow large bubbles, unlike in the slightly older regions
of the Western Loop where there are large \Halpha\ bubbles.  Hence, we
estimate that most of the clusters in the regions surrounding the
Overlap Region are $\la$5 Myr old while the regions in the Western
Loop (regions L, M, T, S, R, and G) have mean ages in the range 5 --
10 Myr.

The cluster ages derived from the $Q$--$Q$ analysis (\S 4.2.2) provide
another age discriminant.  These ages support the age estimates of
$\sim$10 Myr for the clusters in the Western Loop and those bordering
the overlap regions. Roughly 70\% of the bright clusters
in our sample have ages $<$20 Myr.  Unfortunately, because most of these
clusters lie near a loop of the evolutionary tracks in the $Q_1-Q_3$
diagram (i.e., clusters in the log[age] range of 6.5 -- 7.2 have nearly
the same colors), it is difficult to get a finer age discrimination.
In addition, there is a tendency to overestimate the true fraction of
young clusters, since they are more luminous and hence
more easily detected than old clusters.  

The second population of star clusters in The Antennae have ages of
$\sim$100 Myr based on the $Q$--$Q$ analysis.  These objects are found
primarily in the northeastern star formation region (Fig.~5).
Figure 13a shows that roughly 2/3 of the bright clusters in this region
have ages $<$30 Myr and $\sim$1/3 have ages of $\sim$100 Myr. 
The distribution of ages appears to be continuous rather than bimodal. 
The \Halpha\ image (Fig.~4) shows ``streaks'' of recent cluster and
star formation in the northeastern star formation region which are
embedded amongst objects showing no \Halpha\ emission.  
The cluster
candidates with ages $\approx$ 100 Myr 
can be seen in Figure 13a as a number of points centered
around \ub~$= -0.3$, \bv~$= 0.1$. In the $Q_1-Q_3$ diagram they
correspond to the objects around $Q_1 = Q_3 = -0.5$. Figure 13b
shows the corresponding diagrams for the cluster-rich
regions on the PC, where the observations are consistent with
having no clusters with ages $\approx$ 100 Myr.

Interestingly, the clusters of this second population appear to be more
spread out than are the very tight clumps of young clusters (e.g., region
T).  Hence, it appears that the
surface-number density of clusters is related to an age sequence, with
younger regions having a higher cluster density than older regions.

The third population consists of star clusters with ages of $\sim$500 Myr,
based on the $Q$--$Q$ analysis. Unlike the first and second populations,
these clusters appear to have formed in a separate burst, probably
when the long tidal tails were ejected.  Based on dynamical simulations
(Mihos \etal\ 1995), the initial encounter occurred $\sim$200 Myr
ago, in general agreement with our estimate of $\sim$500 Myr since the age
estimates from both the dynamical and the $Q$--$Q$ analysis are probably
only good to about a factor of two. 

We identify the older $\sim$500 Myr population with the initial
encounter, rather than the $\sim$100 Myr population, since the tidal
tails are likely to be the oldest relic of the encounter. The
dynamical models suggest that following the initial encounter the two
galaxies separated, and then later reengaged to form  the current
configuration. This picture is supported by the apparent lack of
clusters with ages $\sim$ 200 Myr, based on the $Q$--$Q$ analysis.

The best examples of the third cluster population are found in the
northwest corner of WF2, where three objects with $M_V \approx -10$
appear off the edge of the main galaxy (objects 5, 6, and 8 in Table
2; identified by squares in Fig.~5b).  Figure 17 shows the
color-magnitude and $Q_1-Q_3$ plots for clusters in this region (which
is called the Northwestern Extension in Fig.~5a).  Note the clump of
about 15 objects at \ub~$= 0.2$ and \bv~$= 0.25$, indicating mean ages
of $\sim$500 Myr. Most of the other clusters in this region appear to
belong to the second population with ages $\approx$ 100 Myr.  The
three bright objects are amazingly similar in color (\bv~$= 0.23$,
0.26, and 0.21) and magnitude ($M_V = -9.86$, $-$9.84, and $-$9.61),
and all have \v16\ values appropriate for clusters on the WF (\v16~$=
2.02$, 2.20, and 2.14). Assuming a present age of $\sim$500 Myr, these
clusters will be $M_V = -7.4$ after 14.5 Gyr, typical of old globular
clusters.  Note that these three clusters appear to be part of a loop
which connects back with the Western Loop around region T.
Unfortunately, most of this loop lies to the northwest of the PC and
is not covered by our WFPC2 images. The reconnection around region T
appears to consist mainly of diffuse light from the original disk
(referred to as the diffuse envelope of NGC 4038 in Paper I, see Malin
1992), although Figure 5b also shows a handful of intermediate-age
clusters at the base of the loop (i.e., objects 4, 7, and 9 from Table
2 and Figure 5b; note the very tight range in magnitudes with objects
4 - 9 all being associated with this extended northwestern loop).

What are the ages of the clusters formed in the long tidal tails?
Unfortunately, our images cover only a small portion of the southern
tail in the southeastern corner of WF3.  We find about a dozen objects,
most of them showing \Halpha\ emission that indicates ongoing star
formation at the base of the tail.  However, the only object bright enough
to definitely be a cluster (number 12 in Table 2 with $M_V = -9.48$; see
Fig.~5b)  has an age of $\sim$500 Myr,
whence it was probably formed at the same time as the clusters in the
northwestern extension, supporting our interpretation that this
population was produced in the initial encounter that formed the
tidal tails.
There are about 30 other good candidates for
intermediate-age clusters [i.e., $M_V < -9$ and $\log(age) = 8.4$ -- 9.0] 
the brightest of which are listed in Table 2 and marked in Figure 5b.

The fourth population consists of old globular clusters from the
original progenitor galaxies.  Roughly 15 candidates were first
identified based on their appearance in Figure 3 (i.e., slightly red
with magnitudes in the range $M_V = -8$ to $-$11). Most of these were
found around NGC 4039, where the background is lower.  A check of their
positions in the $Q$--$Q$ plot showed that 11 of these objects do indeed
have ages of $\sim$10 Gyr.   These 11 objects are included in Figure 5b
and Table 3.  Taking into account the facts that (1) we are only seeing
the bright end of the distribution (which generally peaks around
$M_V = -7.2$), (2) we can only identify the objects in regions of low
background (roughly 1/4 of the field), and (3) globular clusters tend
to crowd around the centers of galaxies, we estimate that the total number
of old globular clusters in The Antennae is at least an order of
magnitude larger than our current sample of 11. 

\bigskip

\centerline{\it 4.3.2 LFs vs. Age}

In this section we compare the LFs for the various populations
of clusters in The Antennae.  Since the vast majority of the clusters
belong to the youngest population (i.e., $<$20 Myr), the LFs discussed
earlier in the present paper are most representative of this population. 

Figure 18 shows a comparison of the LFs for the cluster-rich regions
on the PC ($<$10 Myr old population) and for the fraction of the
clusters with ages in the range 30 -- 160 Myr, based on the $Q$--$Q$
analysis. The latter clusters are found primarily in the northeastern
star formation region, as discussed in the previous section.  The two
LFs are normalized at $M_V = -7.4$ to facilitate the comparison.  We
find a shift of about 1 mag at the bright end of the LF,
which---scaled by the Bruzual-Charlot models and assuming a mean age
of 10 Myr for the younger clusters---would be expected if the older
population had a mean age of $\sim$50 Myr, in reasonable agreement
with our age estimate.  In addition, the bend in the LF appears to
occur at a slightly fainter absolute magnitude for the $\sim$100 Myr
population.

Because of the small number of objects, it is not possible to derive
a meaningful LF for the population of old globular clusters.  However, we note
that the mean absolute magnitudes of the brightest few clusters
from Table 2 are
$M_V \approx -9$, typical of a population of
old globular clusters.

\bigskip

\centerline{\it 4.4 Cluster Sizes}

\centerline{\it 4.4.1 Effective Radii}

As mentioned in \S 1, our earlier measurements of the effective
radii of clusters in NGC 4038/4039 resulted in relatively large
values ($\langle R_{\rm eff}\rangle = 18$ pc, Paper~I), which caused
van den Bergh (1995) to question whether the clusters were young
globular clusters or associations.  In addition, Meurer \etal\ (1995) found
that clusters in nearby starburst galaxies had typical effective radii
of 2 -- 3 pc, and cautioned that larger values measured for more distant
galaxies may be due to crowding and insufficient resolution.

The current data are better suited for size measurements than our
Cycle 2 observations for three reasons:
(1) the repaired optics provide better PSFs, (2) the PC provides a
better sampling of the PSF, and (3) subpixel dithering further improves
the sampling. 

Two different techniques have been used to estimate the effective
radii \reff\ of clusters.  The first method models the observed
profile with a King (1966) model, taking into account the spatial
variations of the PSF across the chip and the different locations of
objects relative to pixel centers (see Kundu \& Whitmore 1998 for
details).  The images from the two dithering positions are measured
separately since we have not yet calibrated this technique for
DRIZZLE-combined data.  The estimates from the two images are then
combined into a final value.  This technique uses model PSFs from the
TINY TIM program (Krist 1995) convolved with King models of various
sizes.  We note that the derived effective radii are roughly
independent of the assumed concentration (Kundu \& Whitmore, 1998).
The use of high S/N PSFs from the actual PC image was not possible
because of the lack of bright stars on the PC. However, one bright
star (Star-6 from Paper I) was placed separately on the PC for a 30~s
exposure to check the PSF at the center.  There is good agreement with
the PSF computed by TINY TIM.

Using this first technique we find that objects in star-dominated regions
appear to have a mean \reff~$\approx 2$ pc. This finite radius probably
results from a combination of three different effects. The first is
that the PSF for the longer exposures in our observations may be
slightly broader (perhaps due to jitter and breathing)
than those used to model the PSF in TINY TIM, which are short
exposures.  The second is that in many regions crowding leads to
overestimating the sizes of objects.  And the third is that some of the
faint objects have the right colors to be clusters rather
than stars, whence they are likely to be slightly resolved.  Because
of these effects, we have chosen to make two size estimates designed
to bracket the true cluster sizes.  An upper limit is determined by
taking the measured values of \reff\ at their face value, with no
correction for the fact that objects in star dominated regions appear
to have sizes of $\sim$2 pc. A lower limit is estimated by subtracting
2 pc from the measured \reff\ to normalize objects in the star-dominated
regions to have a mean value of 0 pc. 

Figure 19 shows the resulting distributions of \reff\ for objects in the
star dominated regions (adjusted to 0 pc), for definite clusters in
uncrowded regions of the PC (i.e., objects with $M_V < -9$), and for
definite clusters throughout the PC.  One important conclusion, based
both on a visual inspection of objects with \reff~$> 10$ pc and a
comparison of the histograms of Figure 19, is that essentially all
objects with \reff~$> 10$ pc are affected by crowding (i.e., overlap
with nearby companions), hence substantiating the concerns of Meurer \etal\
(1995).  Virtually all cluster candidates in uncrowded regions have
\reff~$= 0$ -- 10 pc. Using \reff~$= 10$ pc as a cutoff
results in a median value of \reff~$= 4 \pm 1$ pc for the clusters on
the PC, where the quoted uncertainty  represents the spread between
the upper and lower limits discussed above.  If the 10 pc
cutoff is not applied the median value increases to 6 pc.  Thus the
median \reff~$= 4 \pm 1$ pc is only slightly larger than the median
\reff\ of globular clusters in the Milky Way ($\sim$3 pc, see van den
Bergh 1996). Using the same technique for clusters on the WFC chips
results in much larger scatter (\reff~$= 8 \pm 3$ pc) due to the
more severe undersampling, whence our best estimates for \reff\ are
the PC measurements.  Other recent estimates of cluster radii in
mergers are \reff~$\approx 3$ -- 6 pc
(Schweizer \etal\ 1996; Miller \etal\ 1997; Whitmore \etal\ 1997;
Carlson \etal\ 1998). 

Our second technique for measuring \reff\ makes use of the dithered
data on all chips, but at the expense of adopting the simpler method
of using Gaussian cluster profiles and the measured concentration
indices \v16\ to estimate the sizes, based on numerical experiments
(same method as employed in Paper~I).  The advantages are the improved
spatial resolution due to subpixel dithering and the fact that there
are several bright stars on the WFC images to determine the PSF.  In
addition, the existence of old globular clusters on the same WFC
images makes a comparison between young and old clusters more
straightforward.  The mean values of \v16\ measured for 10 of the 11
stars in Table 4 (dropping the object with \v16~$= 2.17$) is
$<$\v16~$>$$= 1.53 \pm 0.03$ mag (mean error).  For the 11 candidate
old globular clusters from Table 3 the mean value is $<$\v16~$>$$=
1.82 \pm 0.03$ mag. For 21 young clusters from Table 1 on the WFCs
(dropping those with values $>$2.5 mag which are generally in crowded
regions) it is $<$\v16~$>$$= 2.05 \pm 0.04$ mag. And for 20
intermediate-age clusters from Table 2 on the WFC (again, dropping
those with a value $>$2.5 mag) it is $<$\v16~$>$$= 2.06 \pm 0.04$ mag.
The corresponding values for the mean effective radii are $\langle
R_{\rm eff}\rangle = 3.0 \pm 0.3$ pc for old globular clusters, $4.6
\pm 0.4$ pc for young clusters, and $4.7 \pm 0.4$ pc for
intermediate-age clusters.  This shows that the clusters are
relatively well resolved and supports our earlier claim that the young
clusters appear to be slightly larger than old globular clusters,
based on the PC measurements.  We also note the good agreement between
our measurements of \reff\ for old globular clusters in NGC 4038/4039
and the values for globular clusters in the Milky Way quoted by van
den Bergh (1996). The good agreement between estimates based on the
simple method and the more sophisticated method using the King models
is also reassuring. The relative independence of \reff\ on the
concentration index, as mentioned above, is probably responsible.

\bigskip

\centerline{\it 4.4.2 Outer Radii}

While the effective radii of the young clusters appear to be only
slightly larger than those of old globular clusters, the outer radii
of a few of the young clusters in NGC 4038/4039 are much larger,
presumably because they have not been whittled away by the tidal
forces of the galaxies yet.  An extreme case is Knot~S (\#405 in Paper
I), which measures over 900 pc in diameter.  For comparison, only two
globular clusters of the Milky Way have tidal radii larger than 200 pc
(the record holder is NGC 5466 with $R_{\rm t}=240$ pc, see Djorgovski
1993), and all globular clusters in M31 measured so far have $R_{\rm
t}<100$ pc (Cohen \& Freeman 1991; Grillmair et al.\ 1996).

Figure~20 shows the surface-brightness profiles of Knot~S (\# 405),
its neighbor Cluster \#430, and the $\sim$500 Myr old Cluster \#225
for comparison.  Knot~S and Cluster \#430 are highly luminous and lie
in a relatively uncrowded region imaged by the PC.  Their profiles
were derived by a combination of aperture photometry near the center
and multi-object photometry (Lauer 1988) further out.  Both Knot~S and
\#430 feature nearly pure power-law envelopes.  Hence, we are only
able to set upper and lower limits for the values of the core radius
$R_{\rm c}$ and the tidal radius $R_{\rm t}$, respectively, due to the
limited spatial resolution near the center and the lack of a clear
tidal cutoff in the outskirts.  The values are ($R_{\rm c}$, $R_{\rm
t}$)~= ($<$4.2 pc, $>$450 pc) for Knot~S and ($<$4.6 pc, $>$73 pc) for
Cluster \#430.  Similarly, the values of the King (1966) concentration
index $c\equiv \log(R_{\rm t}/R_{\rm c})$ are $c > 2.03$ for Knot~S
and $c > 1.2$ for Cluster \#430.  The normal range for Milky Way
globulars is $0.5 < c < 2.5$.  Improved values of $c$ will depend on
measuring core radii from either deconvolved or new, higher-resolution
images.  We note in passing that Elson, Fall, and Freeman (1987) found
that most of the young star clusters in the Large Magellanic Cloud, with
ages in the range 8 -- 300 Myr, also have
profiles which are not tidally truncated.

In contrast, the intermediate-age Cluster \#225 (object 4 from table 2) ---which appears very isolated on a relatively
featureless back\-ground---shows a distinct cutoff at $R_{\rm t} \approx 50$
pc, a larger apparent core radius of $R_{\rm c} = 5.6$ pc, and a central
surface brightness lower in $V$ by 5 mag than that of Knot~S.  During
its lifetime of $\sim$500 Myr, this cluster has apparently relaxed near
the center, faded by several magnitudes, and lost its outermost stars
due to tidal stripping.  Yet, its concentration index $c \gtrsim 0.95$ places
it well within the range of Milky Way globulars.

Finally, we note that whereas the detailed surface photometry confirms
the integrated magnitude of \#430 given in Table~1, the apparent
magnitude of Knot~S (object \# 405) is $V = 15.63$, corresponding to $M_V = -15.8$, when
integrated to the limit of $r = 450$ pc. Thus, Knot~S is brighter than
Table~1 would indicate (i.e., the aperture correction is much larger than
the standard correction) and hence Knot~S is a super cluster not only
in size, but also in luminosity.

\bigskip

\centerline{\it 4.5 Ultraviolet Spectroscopy of Knots S and K}

Figure 21 shows the spectra of Knots S and K in flux units, as
described in \S 2.2. 
The spectral features are typical for young ($<$10 Myr) star
clusters. The strongest lines are broad stellar-wind features such as
\siiv\ and \civ, and narrow interstellar lines of lower ionization, like
\cii\ and \siii. Note the double structure in most of the resonance
lines. Strong Galactic halo absorption causes an additional component
at a blueshift of about $-$1600 \kms\ in the velocity frame of The
Antennae.  The only stellar {\it photospheric} features that are
discernible above the noise are those of S~V $\lambda$1502,
Si~III $\lambda$1417, and C~III $\lambda$1427.  

\bigskip

\centerline{\it 4.5.1 UV Spectral Slope  and Cluster Mass Estimates}

Since Knot S and K harbor young star clusters with OB stars that
dominate the UV light, the UV spectral slopes are not sensitive to
stellar population properties, but instead are indicative of reddening
by interstellar dust (Calzetti, Kinney, \& Storchi-Bergmann
1994). Standard spectral synthesis models suggest $F_\lambda \propto
\lambda^\beta$ with $\beta \approx -2.5$ between 1200~\AA\ and 2000~\AA\
for unreddened populations (Leitherer et al. 1999). We measure $\beta
= -2.4 \pm 0.3$ and $\beta = -1.1 \pm 0.3$, for Knots S and K,
respectively.  Using these values, along with
Calzetti's (1997) extragalactic attenuation law, results in estimates
of $E(B-V) = 0.01 \pm 0.04$ for Knot~S and $E(B-V) = 0.12 \pm 0.04$
for Knot~K.  After dereddening, the intrinsic fluxes at 1500~\AA\ of
Knots~S and K become $1.1 \times 10^{-14}$ and $8.9 \times
10^{-15}$~\ergs, respectively. Within the errors, both knots are
equally bright at 1500~\AA, and we adopt $F_{1500} = 1.0 \times
10^{-14}$~\ergs\ which corresponds to a luminosity of $L_{1500}
= 4.4 \times 10^{38}$~erg~s$^{-1}$~\AA$^{-1}$.

It is instructive to compare Knots S and K with other young clusters
observed in the UV with IUE or HST. A few well studied cases are R136
in the LMC, NGC~4214\#1, NGC~1569A, NGC~1705A, and NGC~1741B1, which
have $L_{1500} = 6 \times 10^{37}$, $2 \times 10^{38}$, $3 \times
10^{38}$, $6 \times 10^{38}$, and $6 \times
10^{39}$~erg~s$^{-1}$~\AA$^{-1}$, respectively (Leitherer 1998a).
Hence, two of the brightest Antennae clusters have  UV
luminosities that are an order of magnitude more luminous than the
central cluster of 30~Dor, but are rather unimpressive when compared
with cluster B1 in NGC~1741.

\bigskip

\centerline{\it 4.6.2 Age Estimates from the Stellar-Wind Lines}

We can use the standard stellar-wind line technique to determine the
ages of the clusters in Knots S and K  (e.g., Robert, Leitherer, \&
Heckman 1993; Leitherer et al. 1995).  
This technique utilizes the tight relation between the stellar far-UV radiation
field and the wind density in massive hot stars. The winds are driven by 
momentum transfer from photospheric photons to wind material via absorption
lines. Therefore the strength of lines such as \siiv\ and \civ\ correlates
with the far-UV radiation field, and therefore with the relative proportion
of hot, ionizing stars.
The necessary synthesis models
are taken from the Starburst99 package (Leitherer et al. 1999). Although
a large suite of models was explored, we restrict our discussion to models
in the parameter space favored by our previous discussion: (1) We
assume that clusters contain single-burst stellar populations.
(2) The mass spectrum is parameterized as a
power law with a Salpeter slope between 1 and 100~\Ms. The lower mass
cut-off is unimportant since we are only considering normalized
quantities. (3) The metallicity of the evolution models is solar,
and that of the library stars is about 0.2 dex subsolar, corresponding
to the average interstellar metallicity within a few kpc from the Sun.

If these assumptions are made, the only free model parameter necessary
to match the observed spectra is the age of the population. In
Figure~22 we compare the spectra of Knots K and S with model spectra for
cluster ages 1, 3, 5,
7, and 10~Myr.  The most sensitive age indicator is the \siiv\ line, which
is seen as a P~Cygni profile from the winds of massive O supergiants
in 3~--~5~Myr old populations. No P~Cygni profile is observed at
younger ages, and only a relatively weak blueshifted absorption appears
after 5~Myr (see Leitherer \etal\ 1995). The observed \siiv\ profiles
suggest that Knot~K is 2 -- 4 Myr old, and Knot~S is 6 -- 8 Myr
old. Note that the strong narrow absorption components (e.g., \siii)
are of interstellar origin.  The behavior of the \civ\ line is consistent
with our age estimate. The P~Cygni profile of this line decreases
monotonically with age, as opposed to that of the Si~IV line. 
The \civ\ line is significantly stronger in Knot~K
than in S, supporting a younger age of the former.

These age estimates can be used in conjunction with the sizes of the
evacuated \Halpha\ bubbles (radius $= 2\farcs2 \pm 0\farcs2$ around
Knot~S and $0\farcs85 \pm 0\farcs15$ around Knot~K) to derive
effective outflow velocities, $v_{\rm outflow}$. The results are
$v_{\rm outflow} = 29 \pm 5$ \kms\ for Knot~S and $v_{\rm outflow} =
25 \pm 10$ \kms\ for Knot~K.  

The derived age of $\sim$3~Myr for Knot~K suggests that few
supernova events will have occurred since the onset of the
star-formation episode, whence we expect that the mechanical energy
input into the interstellar medium is dominated by stellar winds.  Is
this enough to produce a global wind as suggested by the \Halpha\
bubbles?  A rough estimate of the energy budget with
the Starburst99 package shows that a 3~Myr old population with 40\% solar
metallicity injects about $8 \times 10^{52}$~erg into the gas, more
than enough to form the observed bubble.

To summarize, based on their UV spectra we estimate that Knot~S is
$7 \pm 1$ Myr old, while Knot~K is $3 \pm 1$ Myr old. 
These ages agree quite well with the age estimates based on the $UBVI$
measurements and the $Q$--$Q$ analysis described in \S4.2.2, which
yielded ages of $\sim$5 Myr for both knots.

\bigskip

\centerline{\it 4.6.3 Kinematics of the Interstellar Medium}

Figure 23 shows an enlarged portion of the spectra of Knots S and K,
centered on the \cii\ line. Two line components are visible in each
knot:  one intrinsic to The Antennae and the
other intrinsic to the Galactic-halo component.  The
lines are offset from the nominal wavelength of 1334.53~\AA\
by about $+$1~\AA, which agrees with the expected
uncertainty in the wavelength zero point due to the target acquisition.

The mean measured line widths (FWHM) of the two components
are 455~\kms\ for the Galactic halo lines and 630~\kms\ for the
lines intrinsic to NGC4038/4039. 
The difference is significant
and can be used to constrain the
kinematics of the interstellar medium in NGC 4038/4039 along the lines
of sight to the two knots.  Other resonance lines (like \siii) show
similar behavior, but their widths and velocities are difficult to
measure due to lower S/N ratios and blending. Taking the Galactic halo
lines as indicators of the GHRS line-spread function (determined by the
instrumental profile and the spatial structure of the stellar light),
we can deconvolve the \cii\ lines to estimate the true widths of the
interstellar components in The Antennae.  Superposition in quadrature,
with the assumption that the Galactic halo line has an intrinsic
FWHM~= 100~\kms\ ; 
suggests a FWHM of ~430~\kms\ for the interstellar components
along the line of sight to Knots~S and K.

This value for the velocity width is larger than that for halo lines
in the Milky Way (i.e., maximum values $\approx$ 200 \kms).
If the observed wavelength of the Galactic halo lines is used to
correct the wavelength zero point, we find that the mean velocities
of the interstellar \cii\ along the line of sights to Knots S and K
are 1706 \kms\ and 1678 \kms, respectively.  These velocities are
somewhat higher than the emission-line velocities from Rubin \etal\ (1970),
which are $1619 \pm 5$ \kms\ and $1628 \pm 4$ \kms, respectively.

\bigskip

\centerline{\bf 5. SUMMARY}

The WFPC2 on the Hubble Space Telescope has been used to obtain $UBVI$
and \Halpha\ images of the prototypical merging galaxies NGC 4038/4039
(``The Antennae''). UV spectra have also been obtained with the GHRS
of two of the brightest star clusters.  Over 14,000 point-like objects
have been identified from the broad-band images.  Based on their colors,
a large fraction of these objects appear to be luminous young stars
($M_V > -8$) formed during the merger.  The number of young star clusters
is estimated to be between $\sim$800 (using the conservative criterion
that only objects brighter than $M_V = -9$ are definite clusters) and
$\sim$8000 (using an estimate based on subtracting all definite stars).

Our main results are as follows.

1. Using a variety of different techniques to decouple the cluster and
stellar LFs (i.e., restricting the sample to $M_V < -9$, isolating
cluster-rich regions, using size information, subtracting off stars
identified via colors), we find that the cluster LF has two power-law
segments and a bend at $M_V \approx -10.4$ ($\approx$ --11.4 after
making a correction for extinction).  For absolute magnitudes brighter
than $M_V \approx -10.4$ the power law is steep and has an exponent of
$\alpha = -2.6 \pm 0.2$, while for the range $-10.4 < M_V < -8.0$ the
power law is flatter with $\alpha = -1.7 \pm 0.2$.    It seems not
feasible at present to determine the cluster LF faintward of $M_V
\approx -8$, where individual stars dominate.  The stellar LF in the
range $-9 < M_V < -6$ appears dominated by young red and blue
supergiants and is much steeper, with $\alpha = -2.9 \pm 0.1$.

2. Assuming a typical age of 10 Myr for the clusters, and 1 mag of
extinction, the apparent bend in the LF corresponds to a mass $\approx
1 \times 10^{5}$ M$_{\odot}$, only slightly lower than the characteristic 
mass
of globular clusters in the Milky Way ($\approx$ $2 \times 10^{5}$
M$_{\odot}$).

3. The clusters are slightly resolved, allowing us to determine their
median effective radii, \reff~$= 4 \pm 1$ pc, similar to or slightly
larger than those of globular clusters in our Galaxy.  However, the
outer radii of some of the clusters are much larger than usual.  The
diameter of Knot S is nearly 1 kpc, presumably because this cluster is
very young ($<$10 Myr) and hence has not yet lost many stars in the
outer envelope by tidal forces.

4.  The ages of young clusters can be estimated by a variety of
methods, including $UBVI$ colors, the presence of \Halpha\ ,
the size of \Halpha\ bubbles, and GHRS spectra. The various
methods give consistent results.  The estimated ages indicate that the
youngest clusters lie around the overlap region ($\la$5 Myr), clusters
in the Western Loop are slightly older (5 -- 10 Myr), and the
northeastern star formation region contains clusters with
ages ranging up to 100 Myr.

5. The LF for the $\sim$100 Myr population is shifted faintward
$\sim$1.0 mag in $V$ relative to that of the population of younger
(0 -- 10 Myr) clusters that dominate over most of the rest of the galaxy.

6. An intermediate-age population of star clusters ($\sim$500 Myr)
has also been identified, with the most obvious members being found in 
the Northwest Extension.  This Extension appears to
be part of a loop that was probably extracted from the galaxy during
the initial encounter responsible for forming the tail. 

7. Eleven old globular clusters with $M_V < -8.0$ have been
identified, primarily around NGC 4039, where there is less confusion
with young clusters.  By extrapolating to fainter magnitudes, and
making a rough completeness correction, we estimate that the total
number of old globular clusters in The Antennae is at least an order
of magnitude larger.

8. Age estimates based on GHRS spectroscopy yield 3 $\pm$ 1 Myr for
Knot K (near the center of NGC 4038) and 7 $\pm$ 1 Myr for Knot S in
the Western Loop, in good agreement with the ages derived from the $UBVI$
colors. 

9. Effective gas-outflow velocities from Knots S and K are
$v_{\rm outflow} = 29 \pm 5$ \kms\ and $v_{\rm outflow} = 25 \pm 10$ \kms,
 respectively, based on the estimated cluster ages and the sizes of
the surrounding \Halpha\ bubbles.
The widths of the interstellar lines indicate gas velocity
dispersions of $\sim$400 \kms\ along the lines of sight, distinctly
larger than those measured in absorption lines due to the Milky Way
halo.

\bigskip
\bigskip

Thanks to Arunav Kundu for his help with the cluster size
measurements.  F.S. and B.W.M. thank Sandra Keiser and Michael Acierno
for their cheerful and dedicated computer and programming support.
F.S. also gratefully acknowledges partial support from the NSF through
Grant AST 95-29263. A special thanks to the referee, Jon Holtzman,
who's insightful comments lead to several important improvements
to the paper.

\clearpage

\parskip 6pt plus 3pt minus 2pt



\clearpage 

\centerline{\bf FIGURE CAPTIONS}

\noindent Figure 1 - Region around Knot G showing the two raw \Halpha\
images along the bottom (shifted by 5.5 pixels in X and 5.4 pixels in
Y), the image using the DRIZZLE software to combine the two images to
the upper left, and the image using the ACOADD software to combine the
two images to the upper right. See text for details.

\noindent Figure 2 - Completeness curves as determined from artificial
star experiments for different background levels on the PC and WF.
The background levels are given in units of DN (Data Number).

\noindent Figure 3 - ``True color'' image of NGC 4038/4039 using $UBVI$
WFPC2 images (i.e., $U+B$ for the blue display, $V$ for the green display, 
and $I$ for
the red display; square root image).

\noindent Figure 4 - \Halpha\ image (i.e., same as Fig.~3, but
using the \Halpha\ image in place of the $I$ image for the red display).

\noindent Figure 5a - Identification image for the various regions
discussed in the paper.  See Rubin \etal\ (1970) or Paper I for original
designations. Figure 5b - Identification image for the objects from
Tables 1 -- 4. Open circles mark young clusters ($<$30 Myr), open
squares intermediate-age clusters ($\sim$500 Myr), and filled
circles old globular clusters.  Plus signs mark foreground stars.

\noindent Figure 6 - ``True color'' enlargements of various regions of
interest in NGC 4038/4039. Regions T and S have been rotated by 90
degrees with respect to Figure 3. Note that the figures have different
scales (see the 2$\arcsec$ bar).

\noindent Figure 7 - \Halpha\ image enlargements of various regions of
interest in NGC 4038/4039.

\noindent Figure 8 - The top row of panels shows plots of the
concentration index \v16\ (a rough measure of cluster size), the \vi\
vs.\ \Mv\ color-magnitude diagram, and the \ub\ vs.\ \vi\ color-color
diagram for objects with $M_V < -9$ (i.e., primarily clusters) in the
cluster-rich regions on the PC (i.e., Knots T, S, R, and \# 13). The
arrow shows the reddening vector.  The middle row of panels shows the
same plots, but for objects with $M_V < -6$ (i.e., stars and
clusters).  The bottom row of panels is for objects with $M_V < -6$ in
cluster-poor regions on the PC (i.e., primarily stars).  Note how the
\v16\ and \ub\ vs.\ \vi\ plots can help separate stars and clusters.

\noindent Figure 9 - Raw LF (lower line) and completeness-corrected (but not reddening
corrected) LFs for objects on all four chips, as well as on the PC alone
(bottom panel).  The 50\% completeness limits are indicated by arrows.
The top panel shows the Number vs.\ \mv\ plot (in which a population of
old globular clusters would have a roughly Gaussian distribution),
while the middle and bottom panels show $\log$(Number)
vs.\ \mv\ plots (where a power law is a straight line).

\noindent Figure 10 - Completeness-corrected LF for the cluster-rich
regions on the PC, as well as the LF for the objects in cluster-poor
regions (primarily stars). The slopes and the location of the
bend as determined by a two-slope power-law fitting function
are included. 

\noindent Figure 11 - Four attempts to separate the LFs of the clusters
and stars.  Panel A shows the total LF for
NGC 4038/4039 (dominated by clusters at the bright end and by stars at
the faint end). Panel B shows the LF for the cluster-rich
regions throughout NGC 4038/4039, along with a version with an attempt
to subtract off the stars (dotted line).  Panel C shows
the LF for objects on the PC that appear to be resolved (i.e., \v16\ $>$
2.5). Panel D shows the cluster candidates based
on the $Q$--$Q$ analysis.  A change in the slope at $M_V \approx -10.4$
is seen in all four histograms. The two-slope power-law fits are shown. 

\noindent Figure 12 - Same data as in Figure 11, but with Schechter-function
fits.

\noindent Figure 13a - \ub\ vs.\ \bv\ plot and the corresponding
$Q_1-Q_3$ plot for a dust-poor region (northeastern star formation
region).  Evolutionary tracks for the BC96 model clusters are
included, along with loci for various types of stars.  The arrow shows
the reddening vector. Small crosses with corresponding labels indicate
the values of log(age) along the Bruzual-Charlot models. As expected,
the reddening from the dust moves the data points off the BC96 tracks.
Figure 13b shows the same plots for the cluster-rich regions on the
PC. Note the clusters with apparent ages $\sim$100 Myr for the
northeastern star formation region. This population is missing for the
cluster-rich regions on the PC.

\noindent Figure 14 - Same as Figure 13, but for objects in the
dust-rich Overlap Region.  
A comparison of the $Q_1-Q_3$ plots in Figures 13 and 14
shows that the two regions both have clusters
with ages $\sim$10 Myr, but a sizable fraction of clusters
with ages $\sim$100 Myr is seen only in the northeastern
star formation region.

\noindent Figure 15 - The dereddened and
completeness-corrected LF for the cluster-rich regions on the PC. This
compares fairly well with the top panel of Figure 10, supporting our
claim that the extinction by dust is relatively small in most of the
cluster-rich regions.  

\noindent Figure 16 - \bv\ vs.\ \vi\ diagram for bright cluster
candidates on the PC. Open circles mark clusters with strong
\Halpha\ emission, while filled circles mark clusters
with weak or no \Halpha\ emission.
Note that the objects with weak \Halpha\ emission fall nicely on the
Bruzual-Charlot model tracks, while objects with strong \Halpha\
emission are offset by about $-$0.2 mag in \vi.

\noindent Figure 17 - \ub\ vs.\ \bv\ and $Q_1-Q_3$ plots for the
bright objects in the Northwestern Extension.  Note the clump of
about 15 objects with apparent mean ages of $\sim$500 Myr (filled circles).

\noindent Figure 18 - A comparison of the dereddened LFs for objects
in the cluster-rich regions on the PC ($\sim$10 Myr) and for the
fraction of the clusters with ages in the range 30 -- 160 Myr, based
on the $Q$--$Q$ analysis. The latter clusters are found primarily in
the northeastern star formation region.  The two LFs are normalized at
$M_V = -7.4$ to facilitate the comparison.  We find a shift in the LF
of about 1 mag at the bright end.

\noindent Figure 19 - Distributions of effective radii \reff\ for objects
in star-dominated regions on
the PC (normalized to 0.0 pc, see text), bright cluster candidates in
uncrowded regions, and bright cluster candidates in crowded
regions.  This figure suggests that most of the objects with apparent
\reff~$> 10$ pc are due to superpositions in crowded regions.

\noindent Figure 20 - Surface-brightness profiles for Knot~S (i.e.,
Cluster \#405), its neighbor Cluster \#430, and the $\sim$500 Myr old
Cluster \#225, all measured from the dithered PC image. No attempt at
deconvolution has been made. Note the power-law nature of the profiles
for Knot~S and \#430, the large extent of the envelope of Knot~S (see
parsec scale at top), and the distinct cutoff at $R_{\rm t} \approx
50$ pc for Cluster \# 225. The vertical scales for the three plots have
not been altered; the central surface brightness of the older Cluster
\# 225 is lower in V by 5 magnitudes.

\noindent Figure 21 - GHRS spectra of Knots S and K obtained through
the Large Science Aperture ($1\farcs7 \times 1\farcs7$). The wavelength
scale is in the restframe of NGC~4038/4039.

\noindent Figure 22 -Comparison between the observed GHRS spectra
(thick lines) and model spectra based on evolutionary population synthesis
(thin lines). The
continuum level of each spectrum is at unity, with offsets of 0 to 7
added to each spectrum from the bottom to the top, respectively.  The
five models are for a single-star population with Salpeter IMF between
1 and 100~\Ms, with ages marked.

\noindent Figure 23 - Blowup from Figure~21 showing the \cii\ lines in Knot~S
and K. The dashed vertical lines mark the nominal rest
wavelength of C~II at 1334.53~\AA\ and the wavelength of 1327.23~\AA,
corresponding to a blueshift of --1642~\kms. The blue components
represent absorption lines due to the Milky Way halo.  Measured line
widths (FWHM in \kms) are given for all lines.

\end{large}			

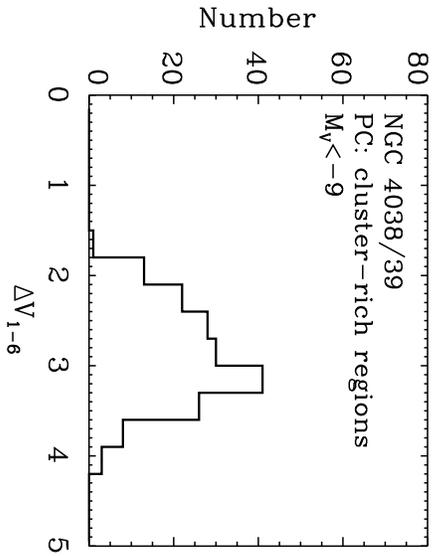
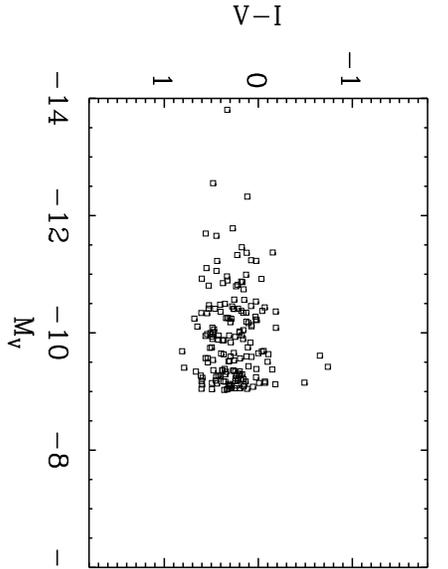
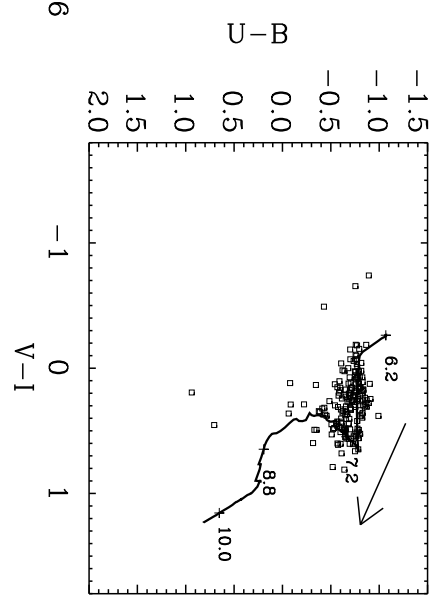
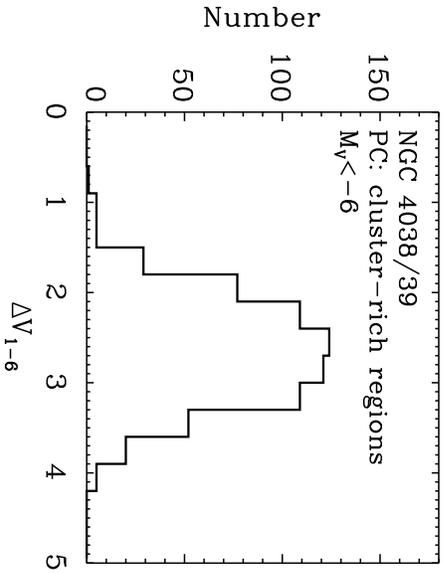
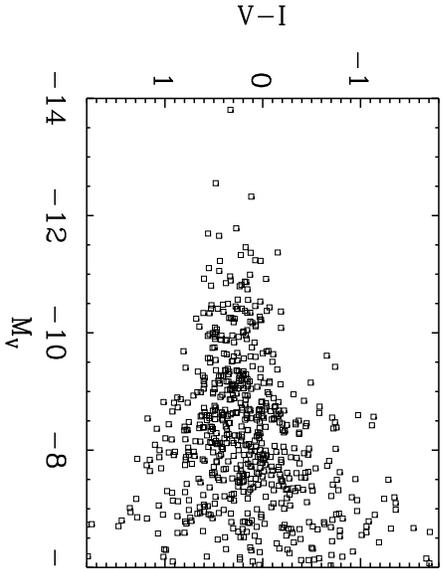
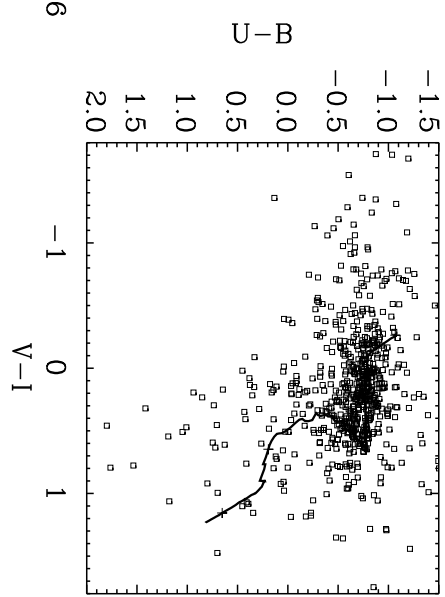
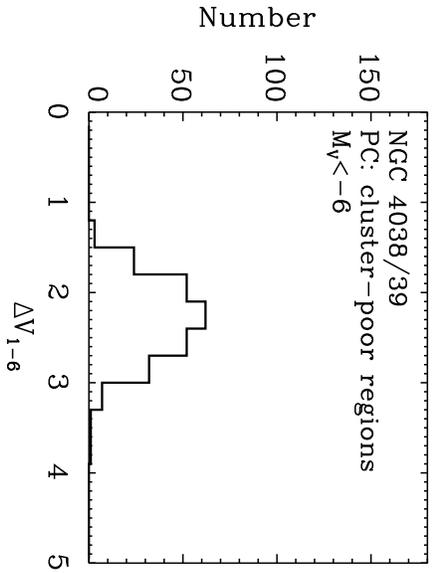
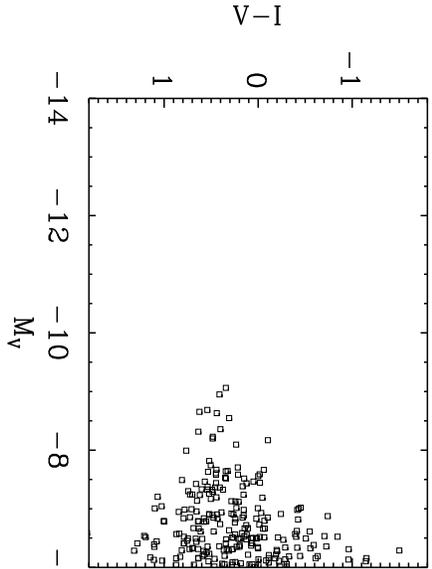
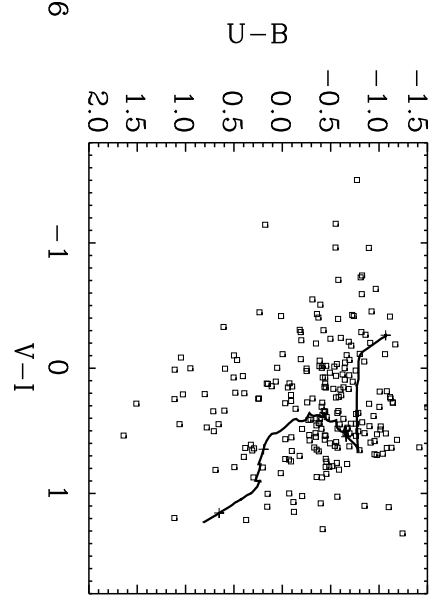
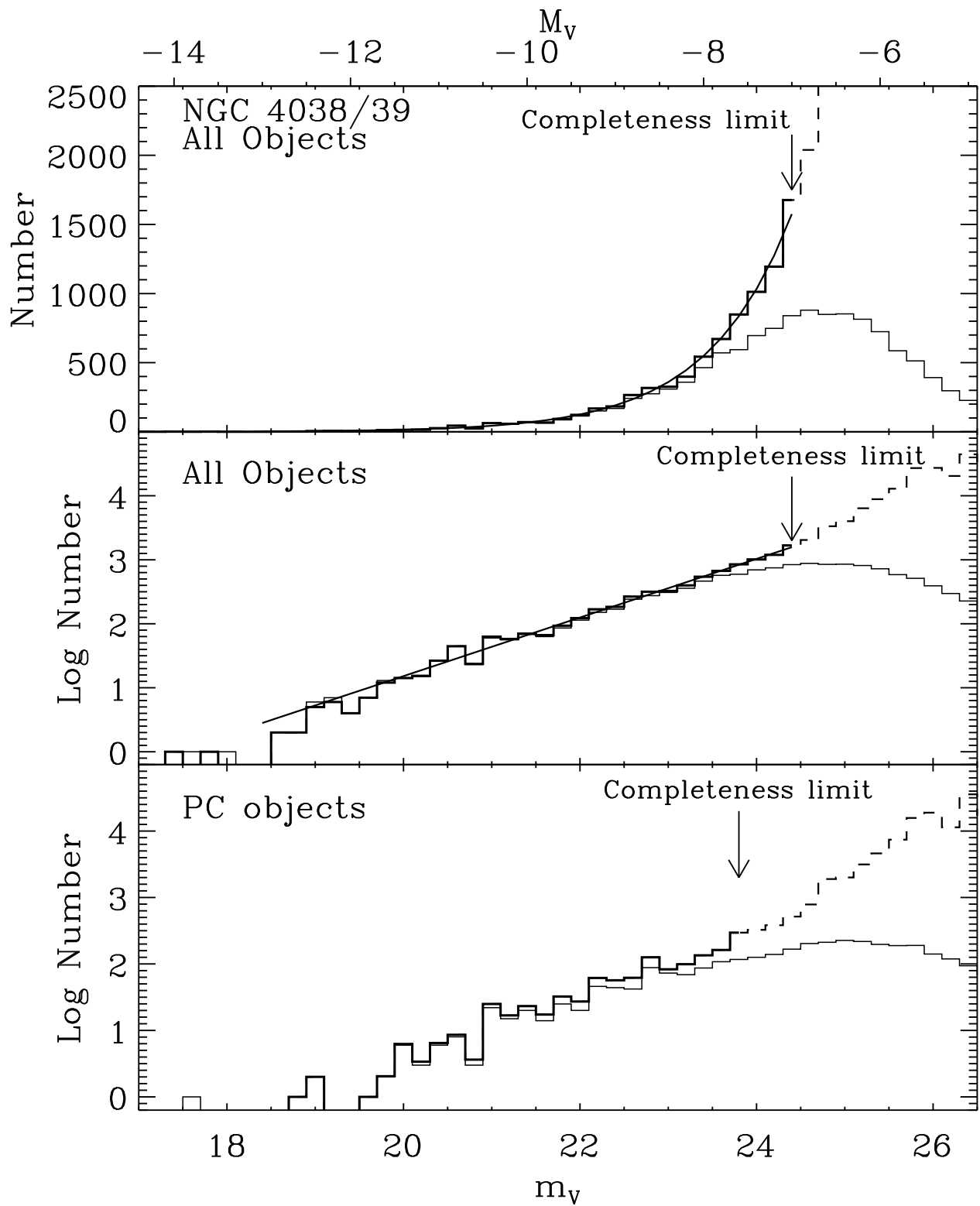
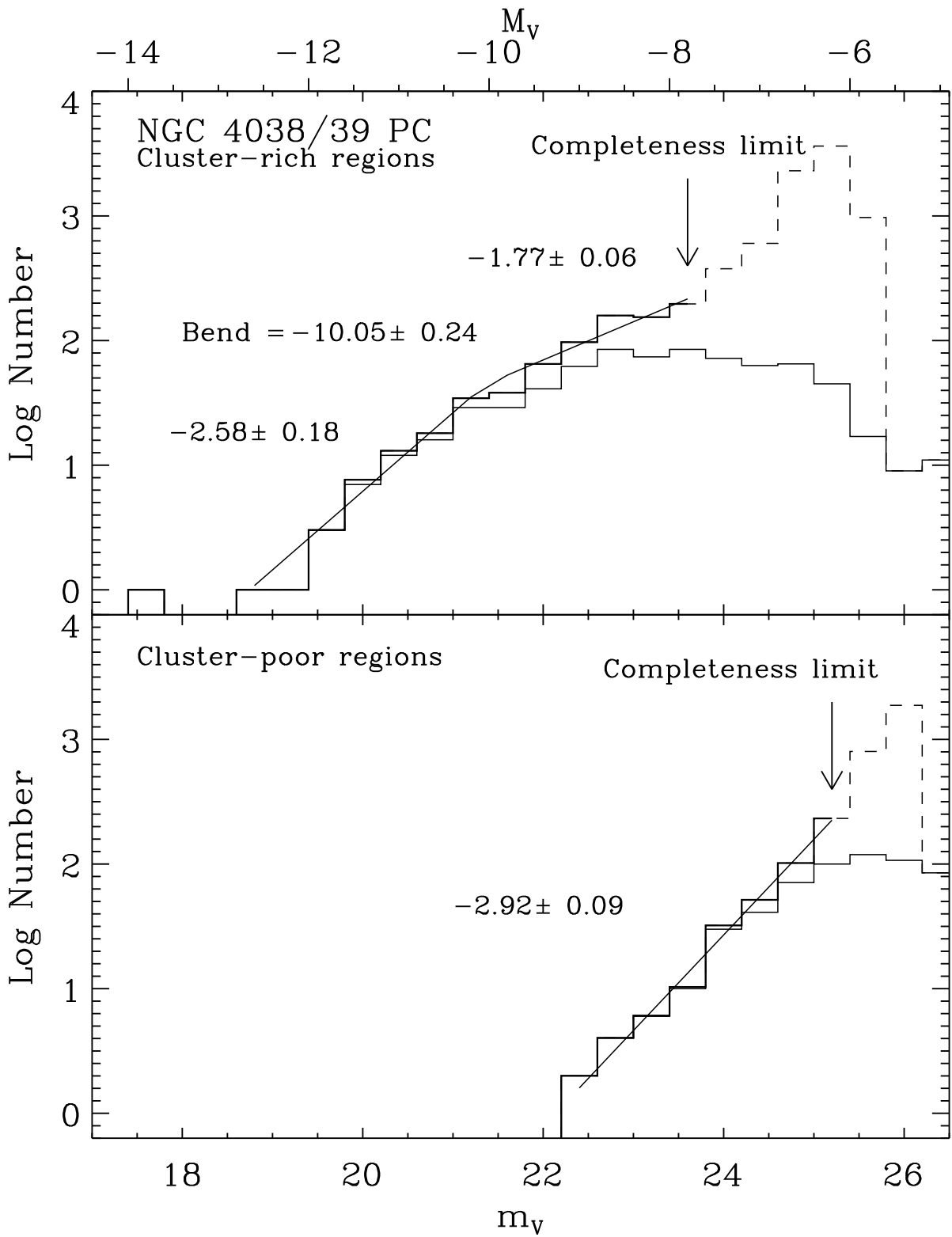
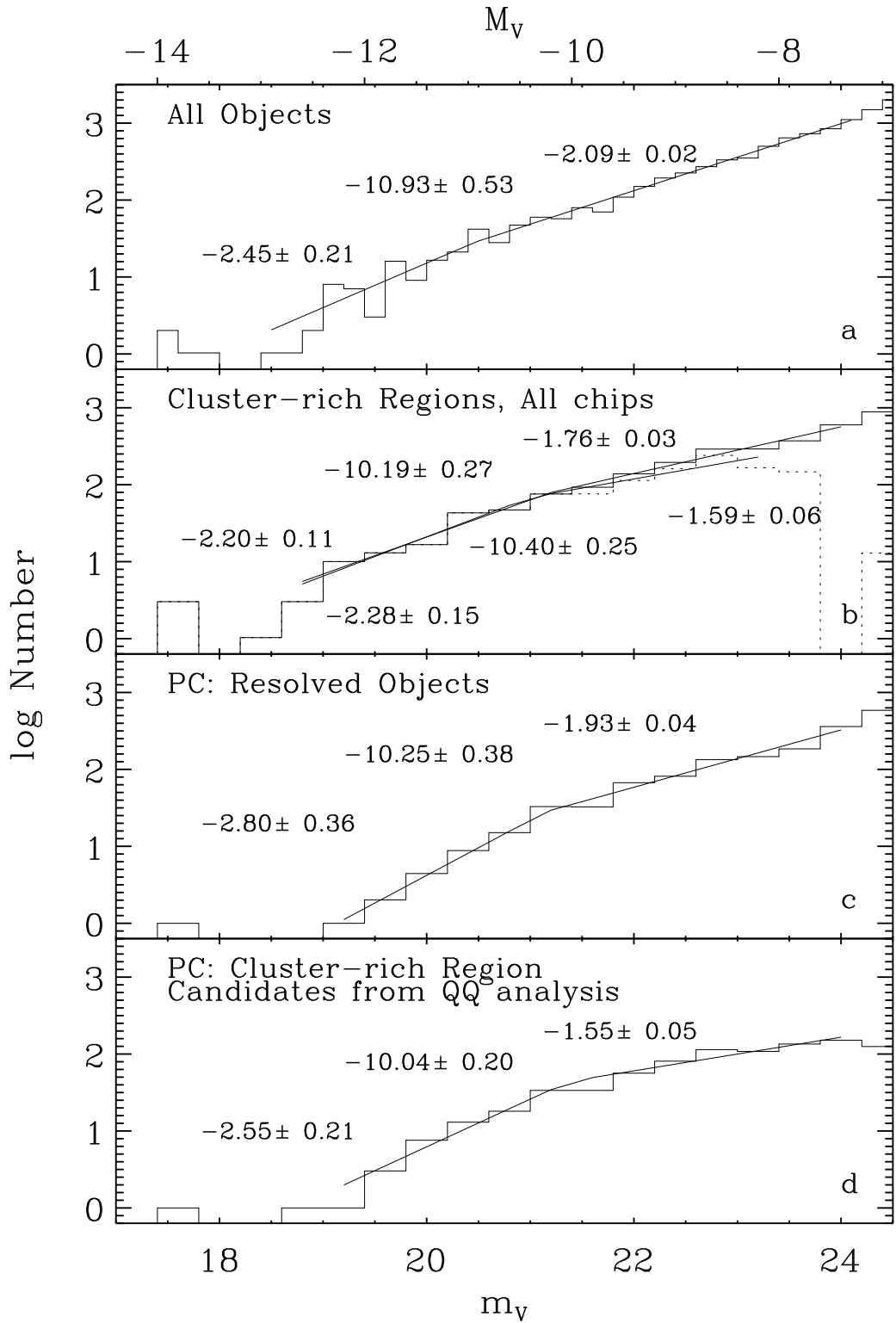
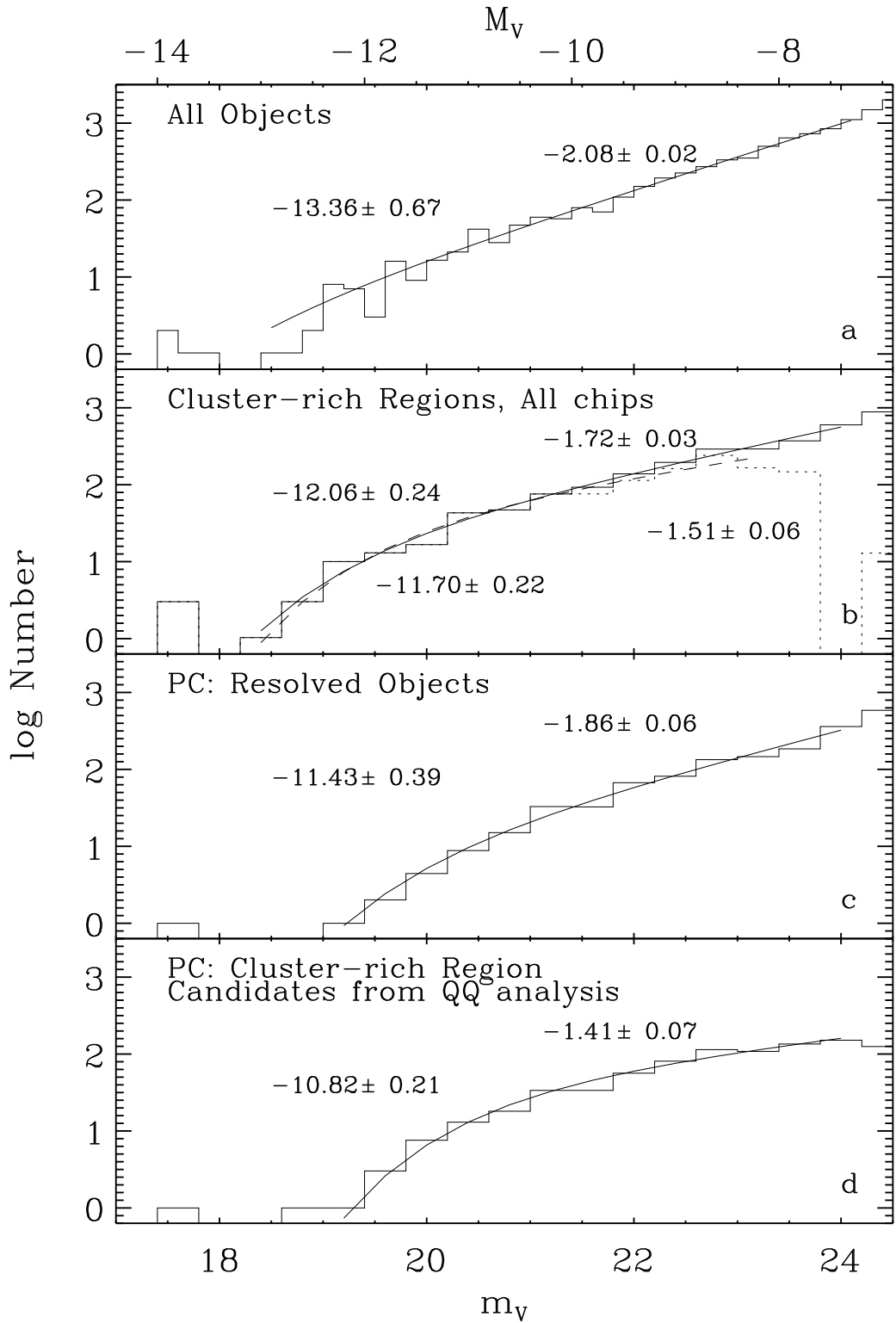
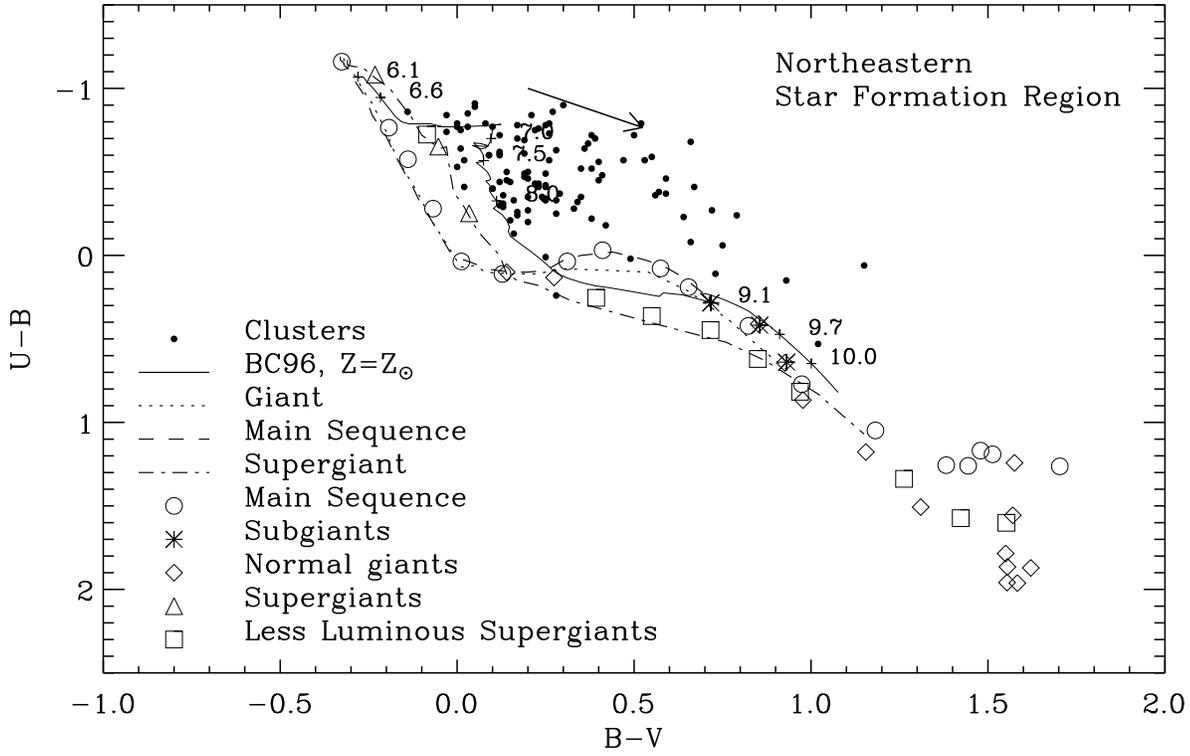
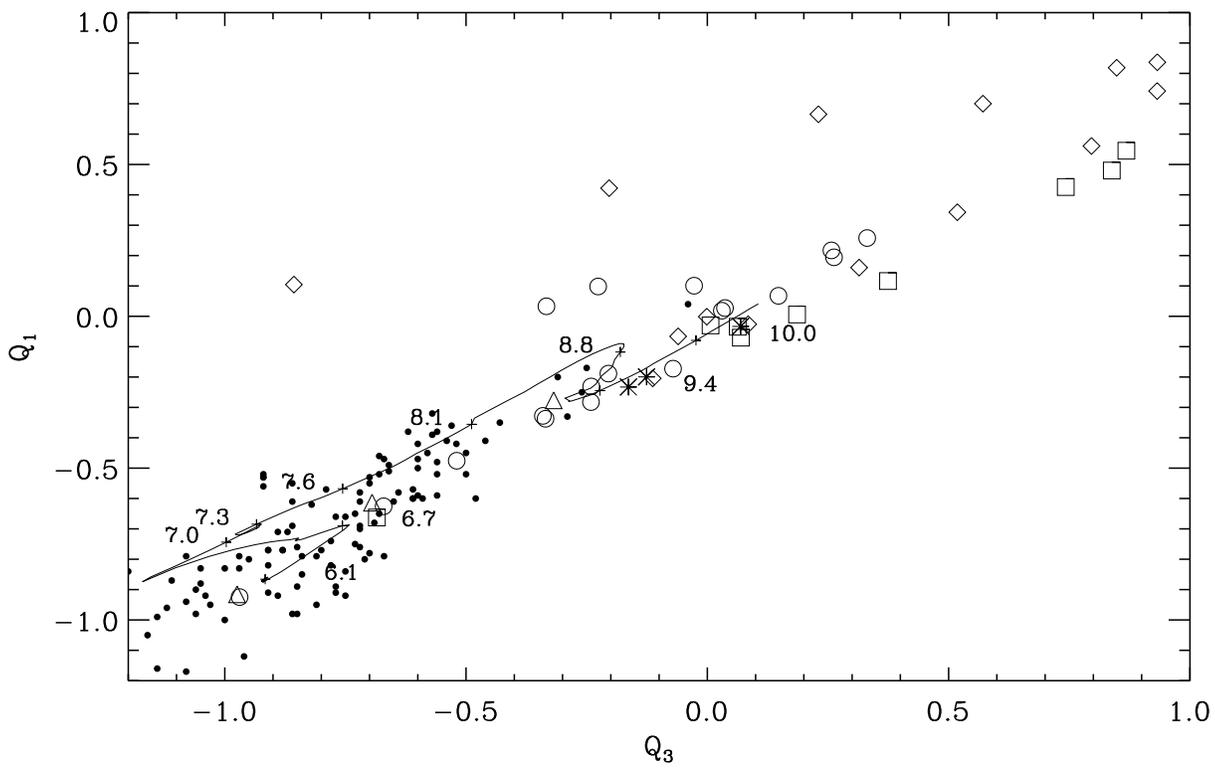
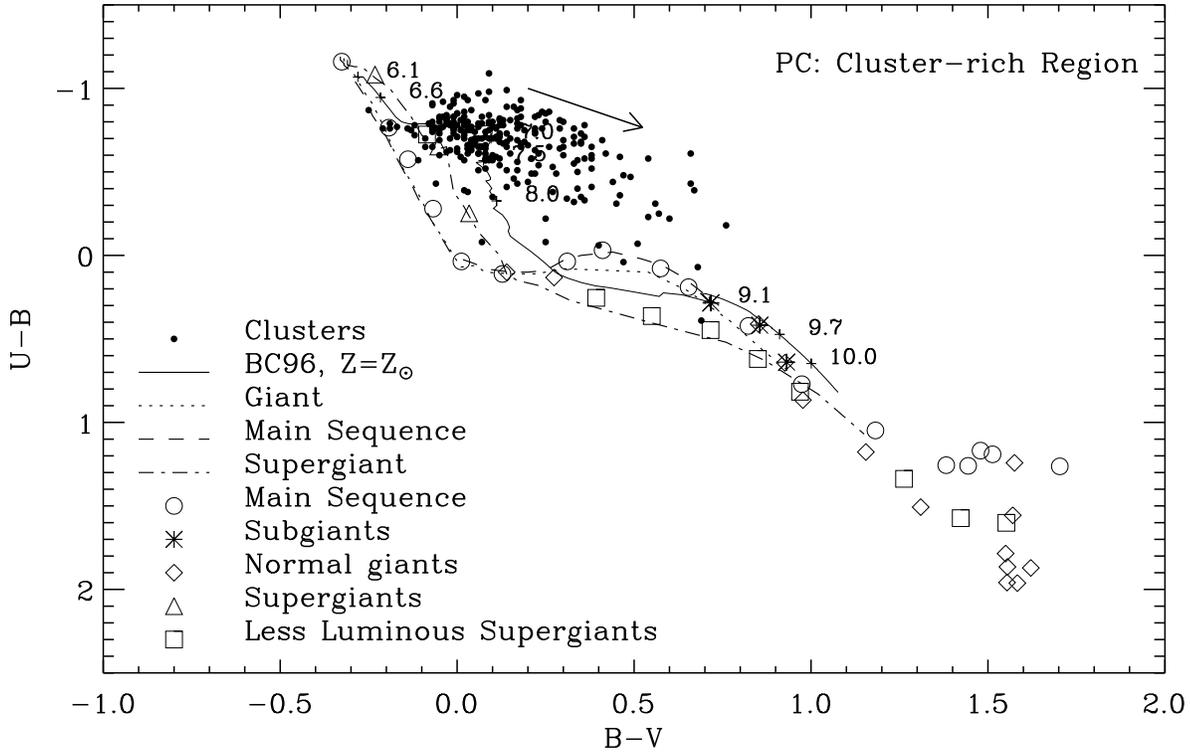
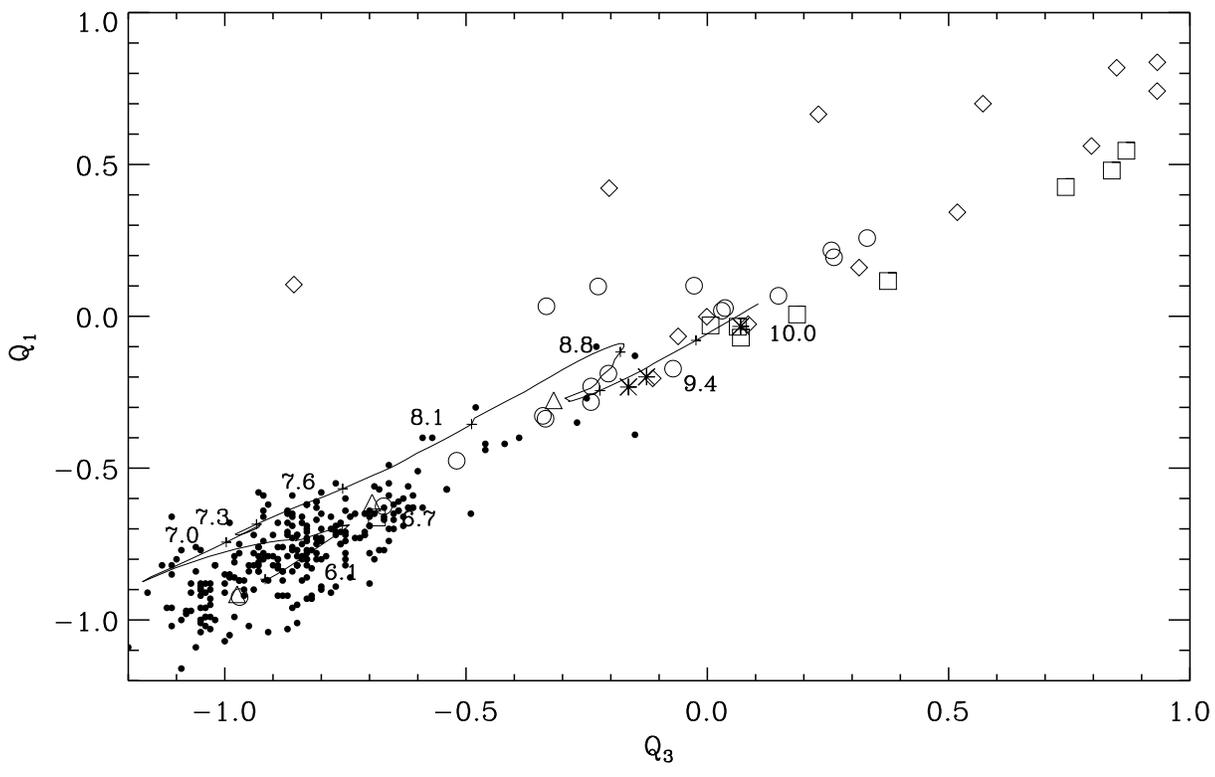
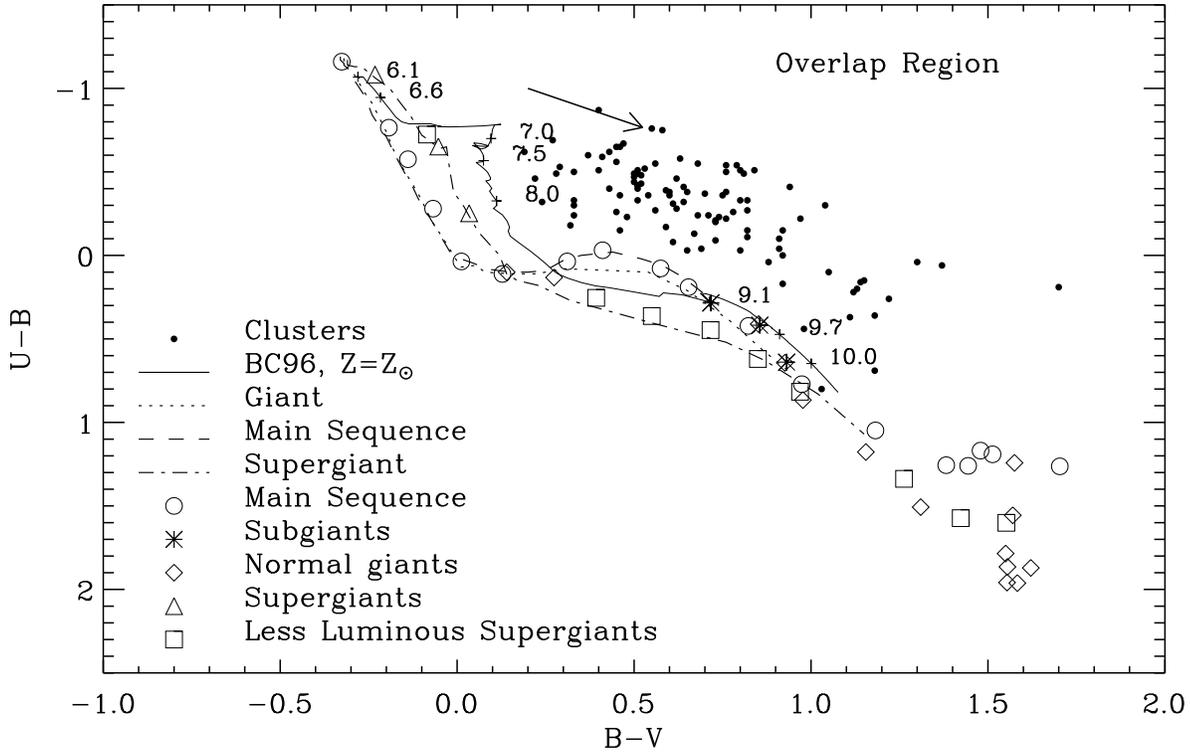
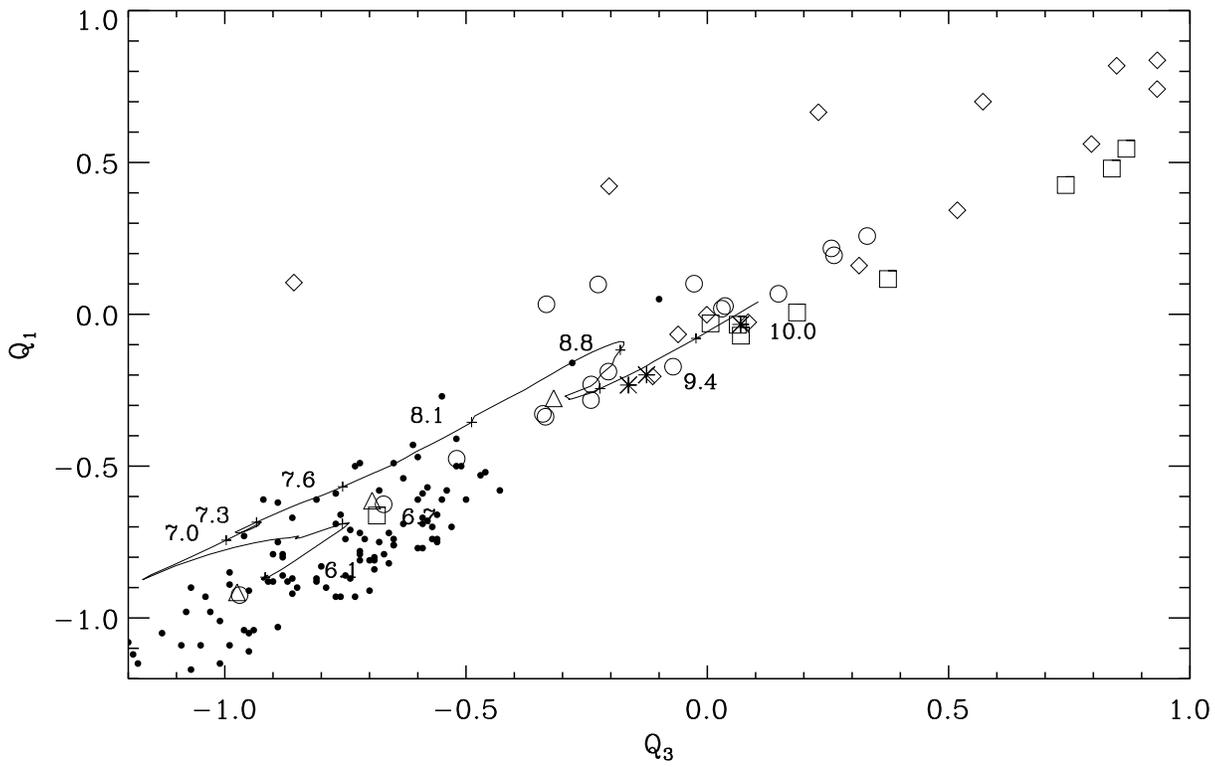
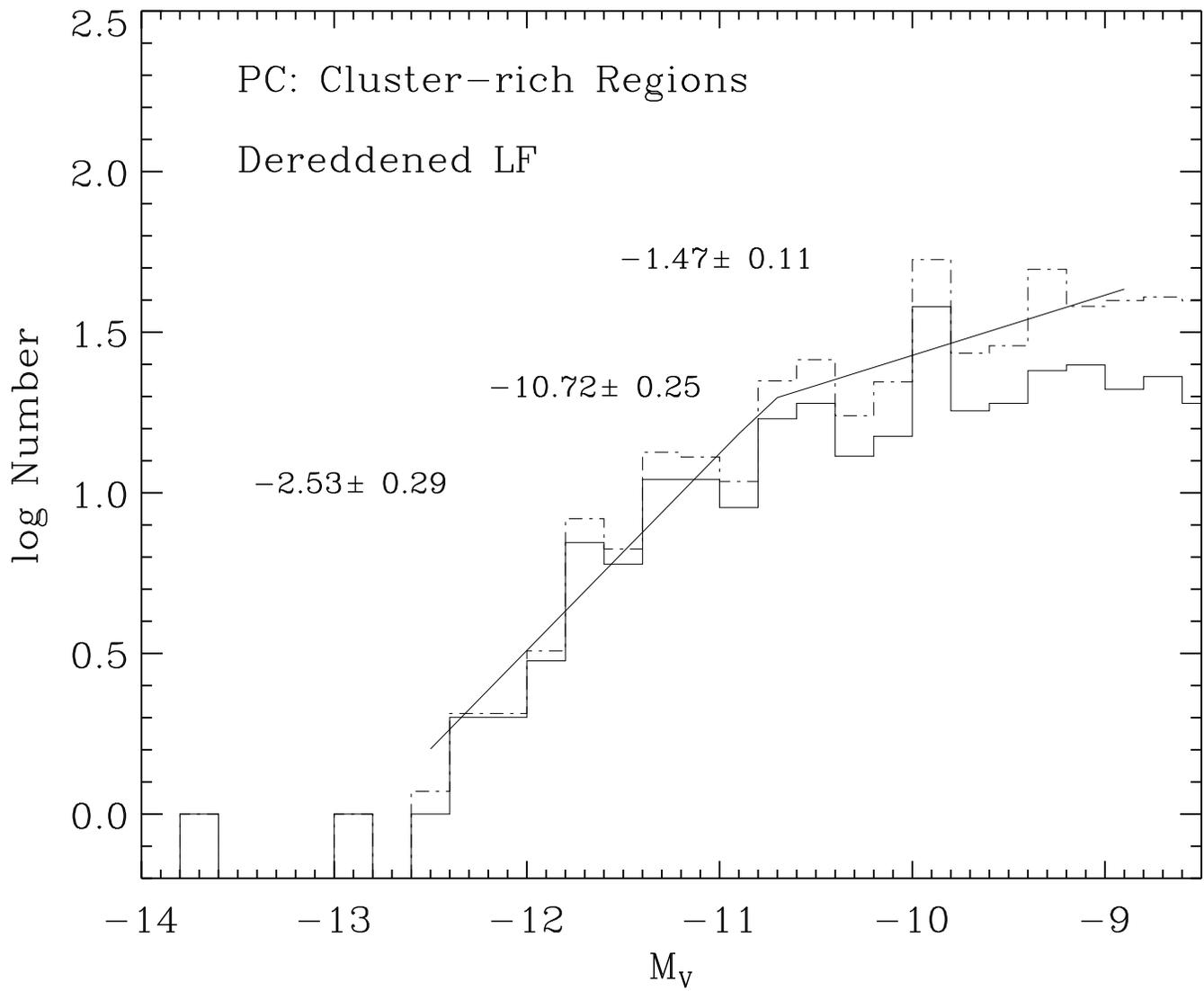
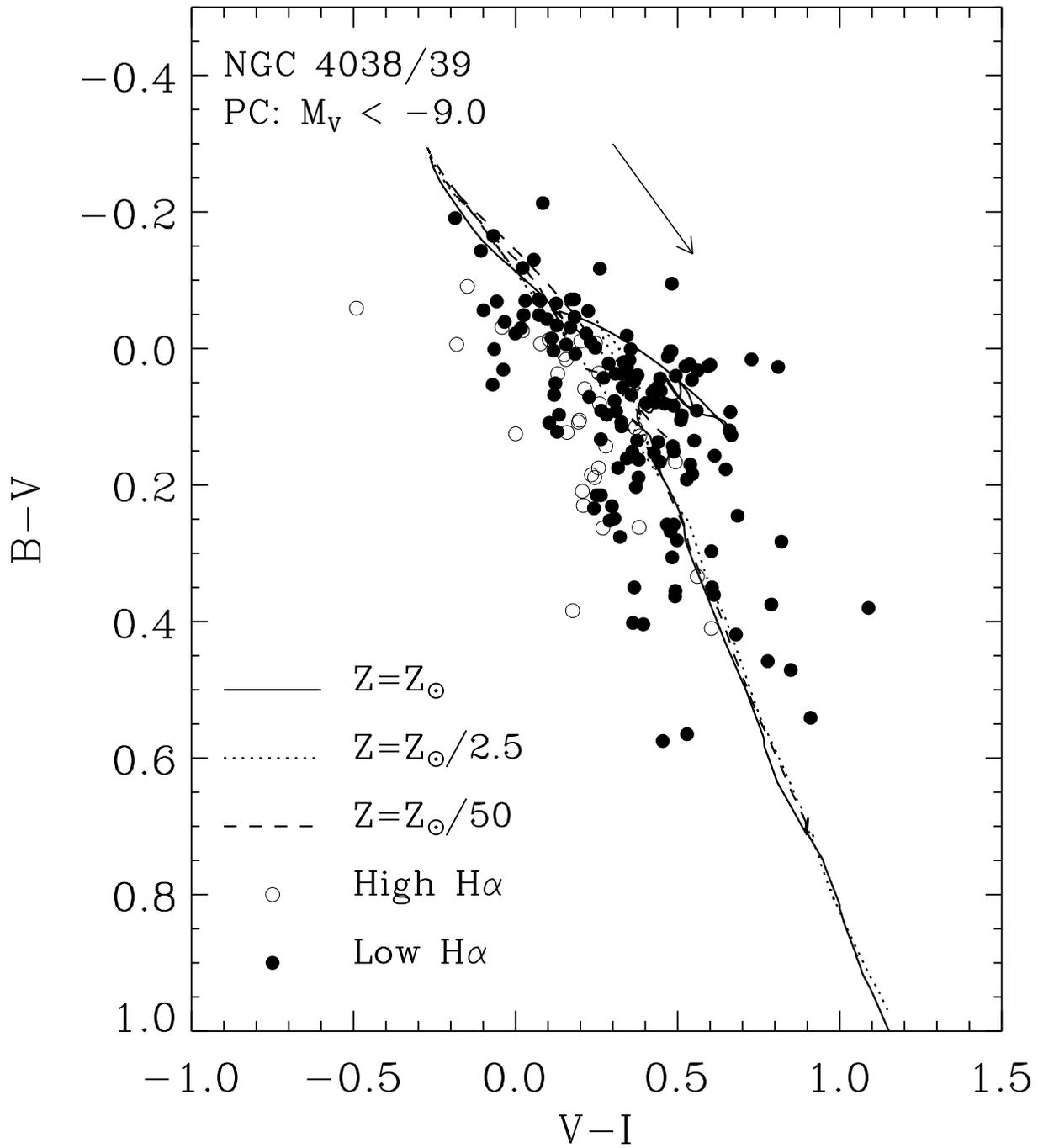
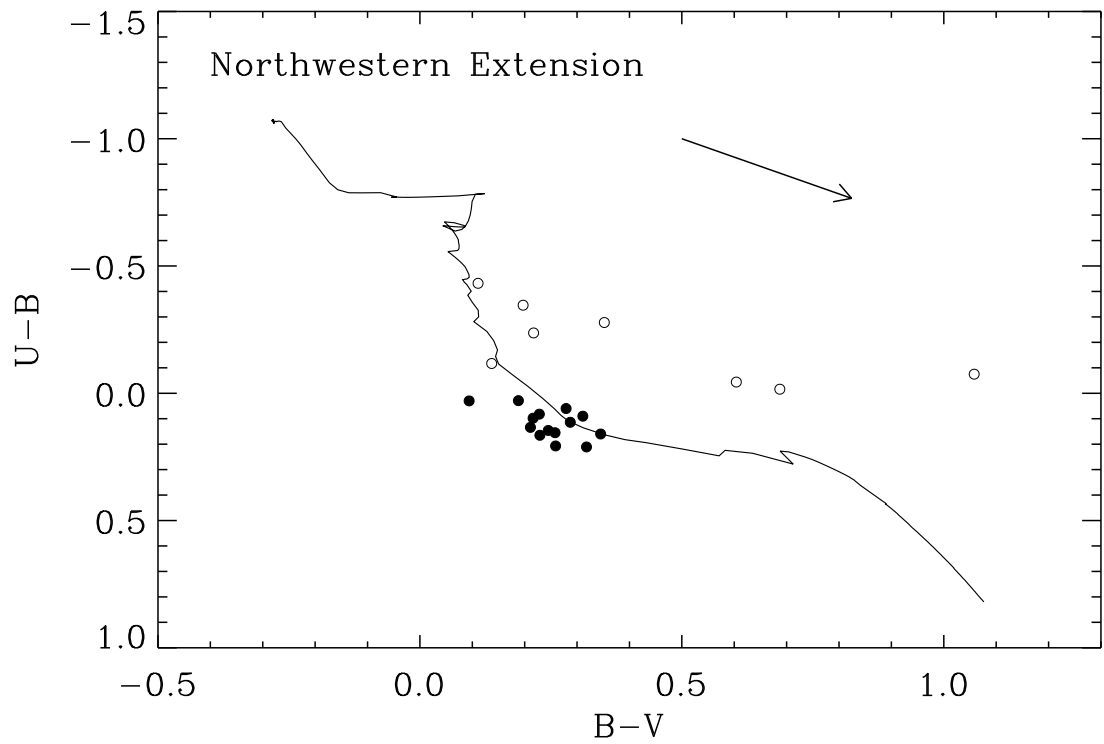
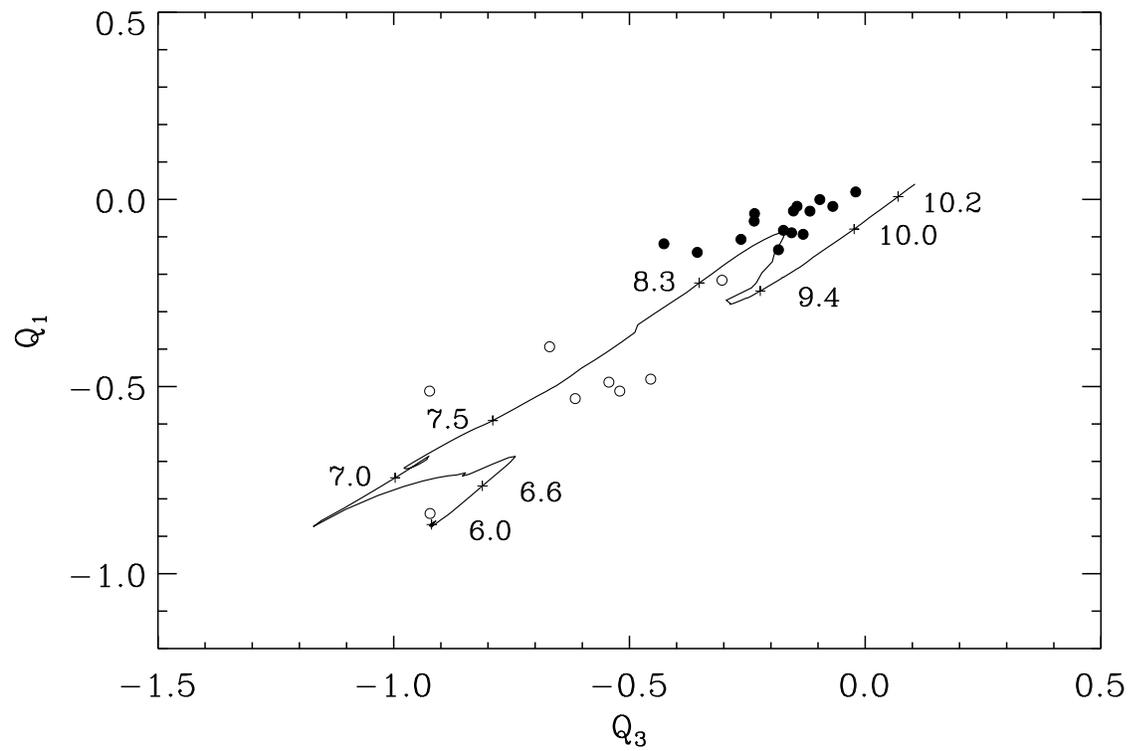
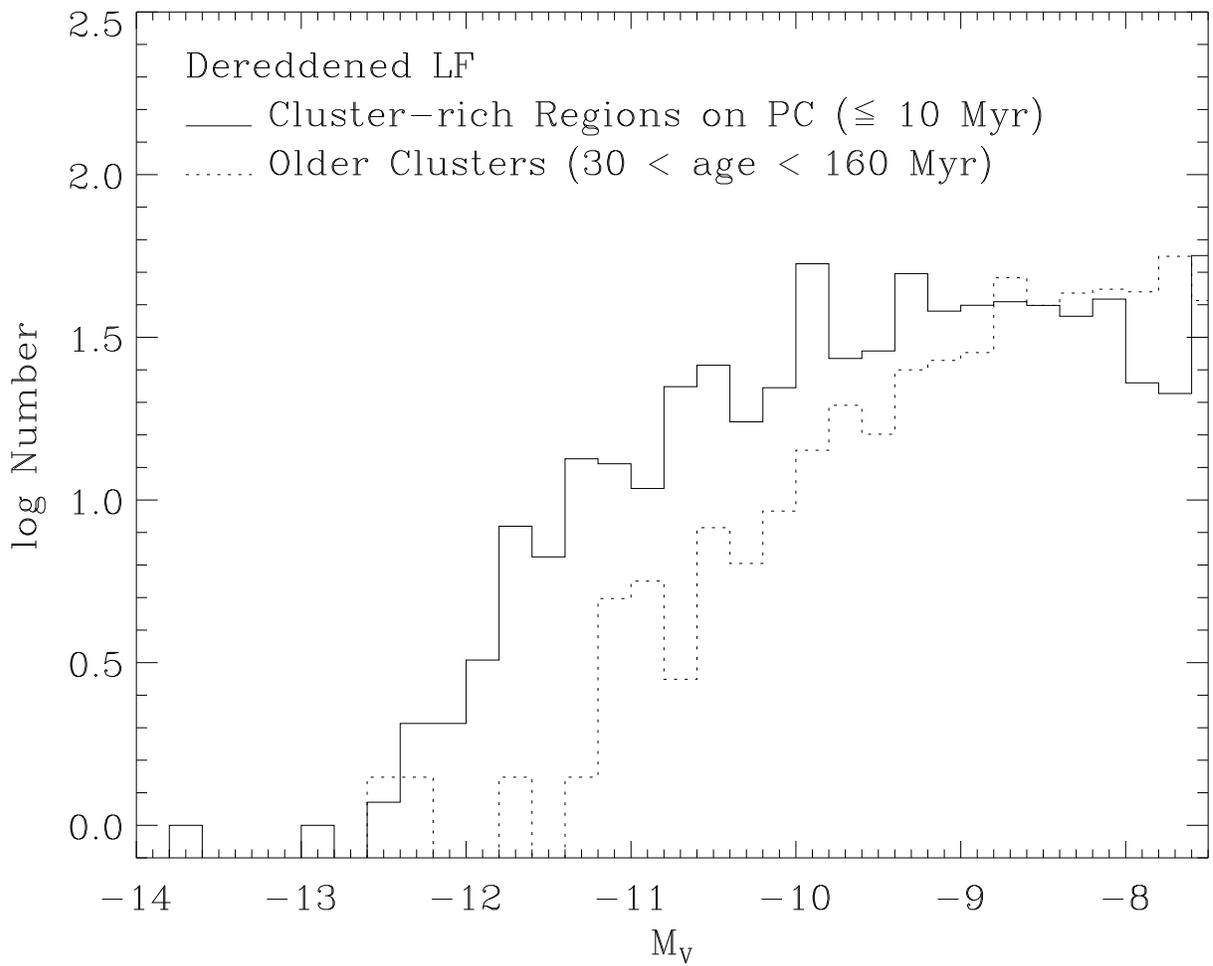
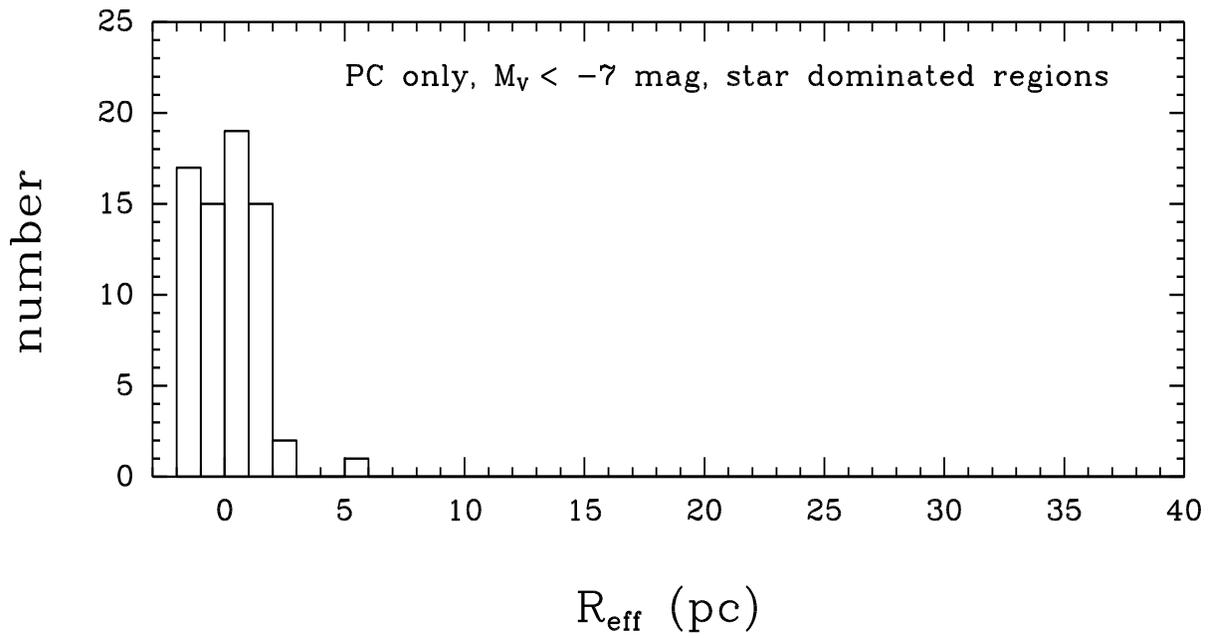
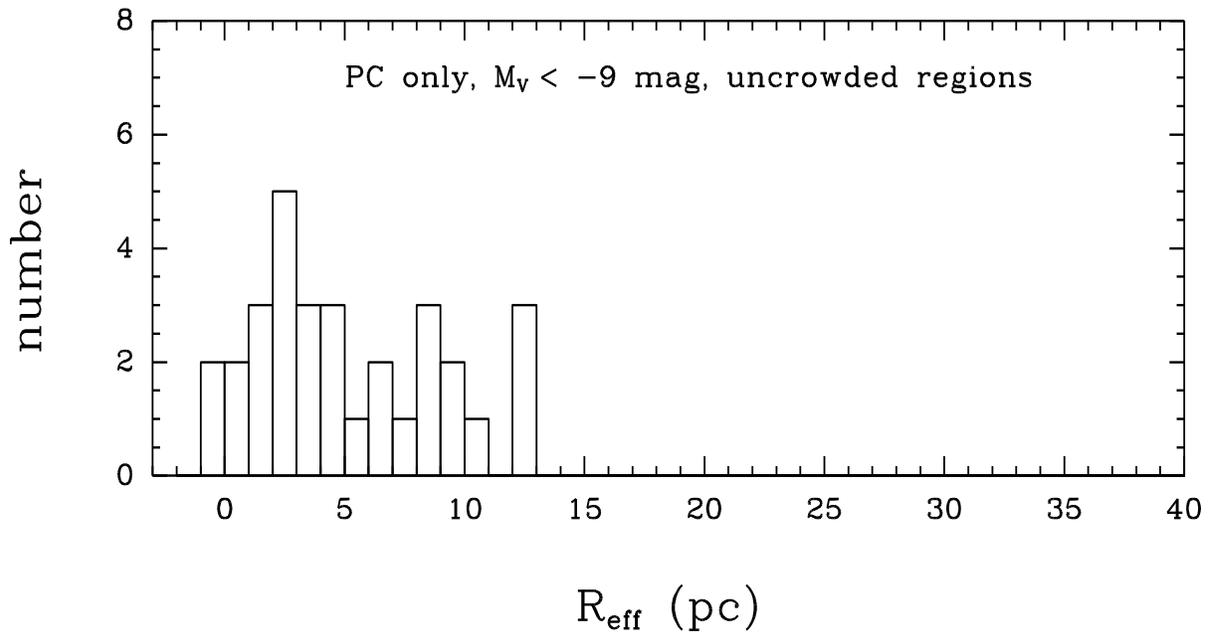
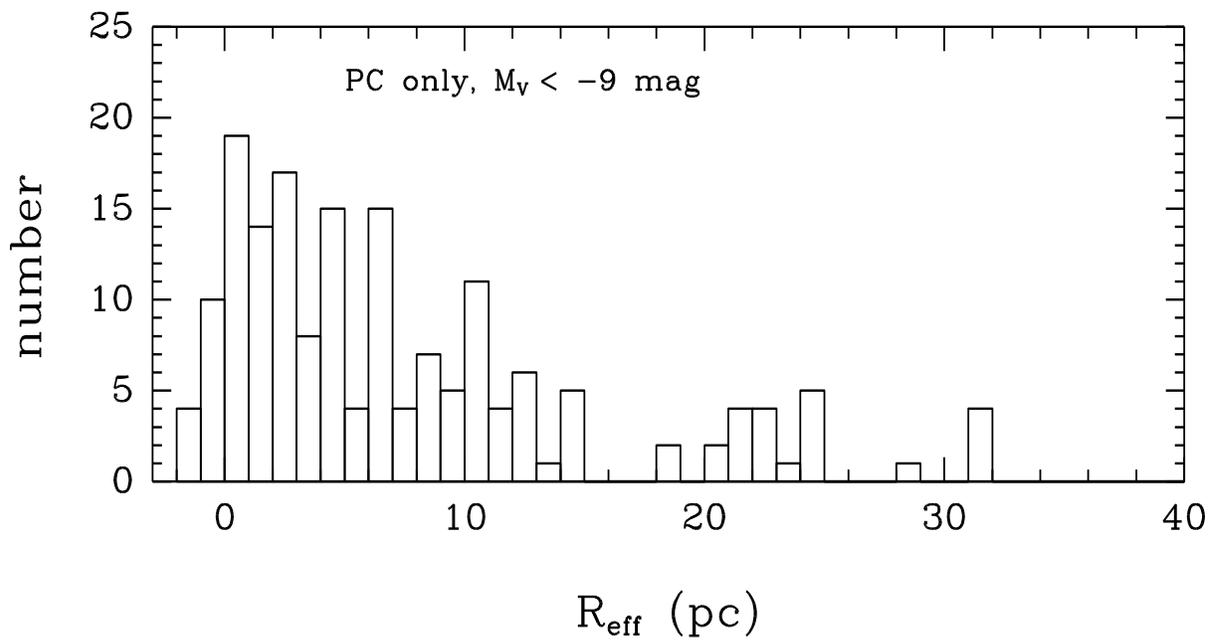
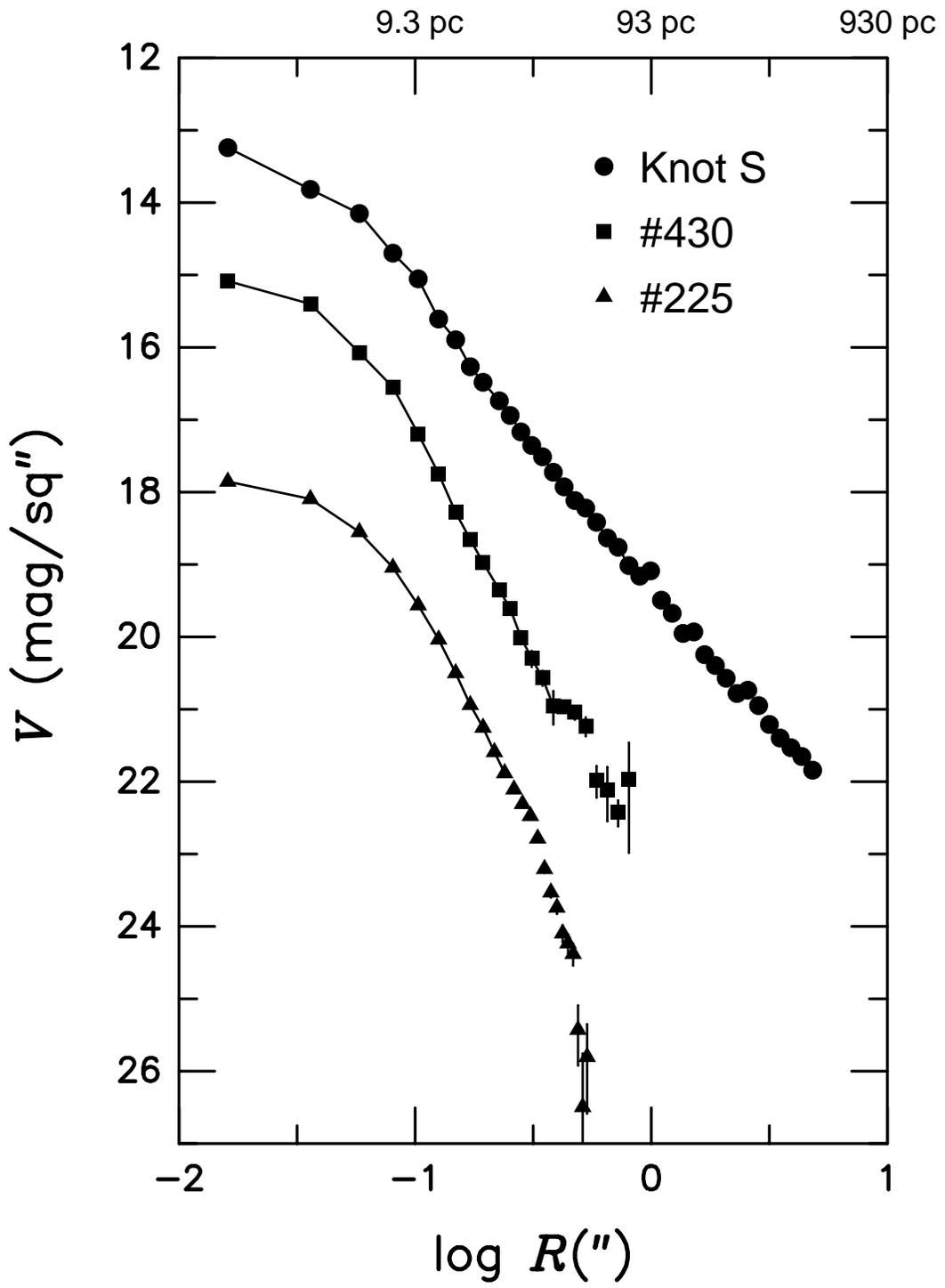
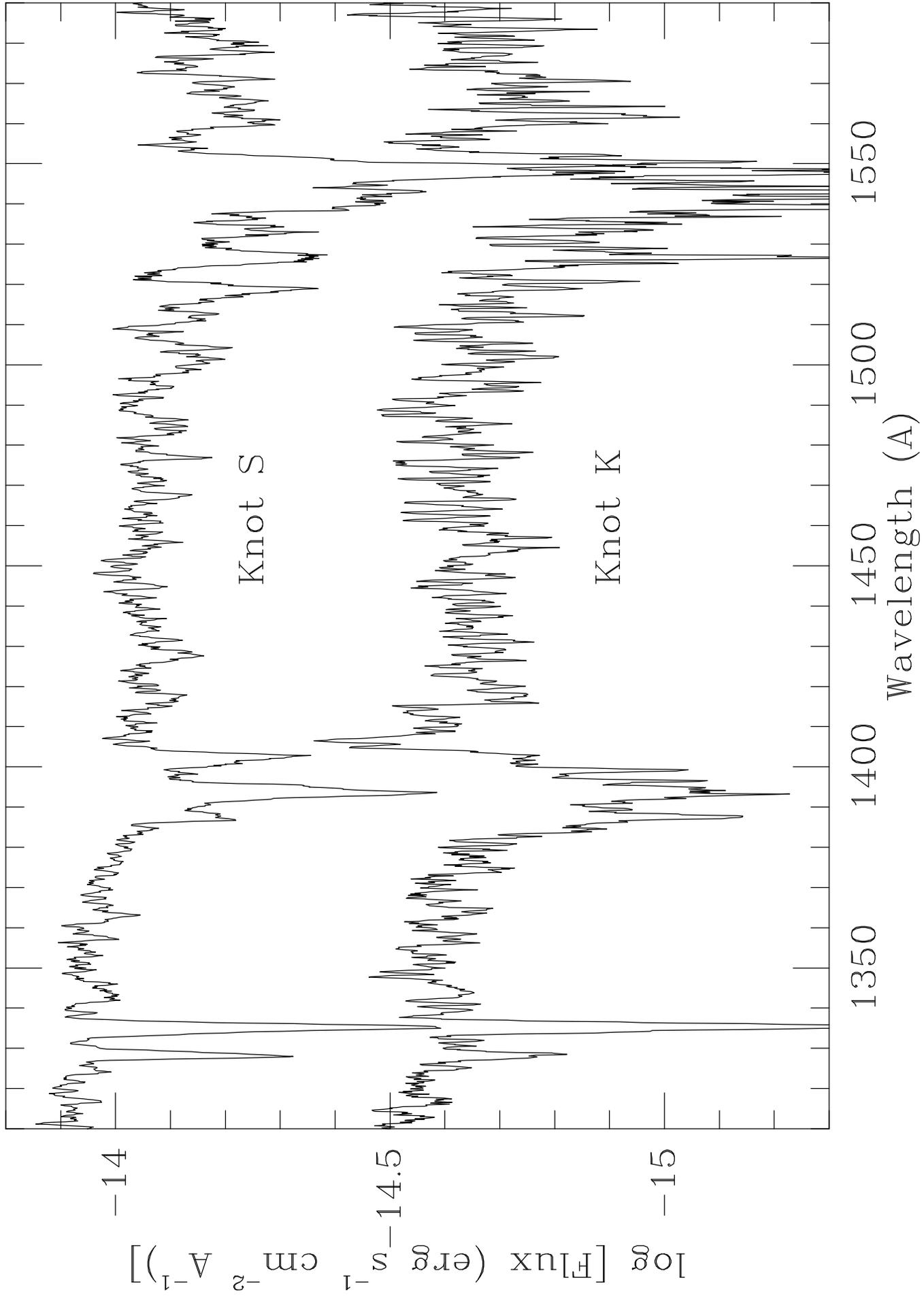
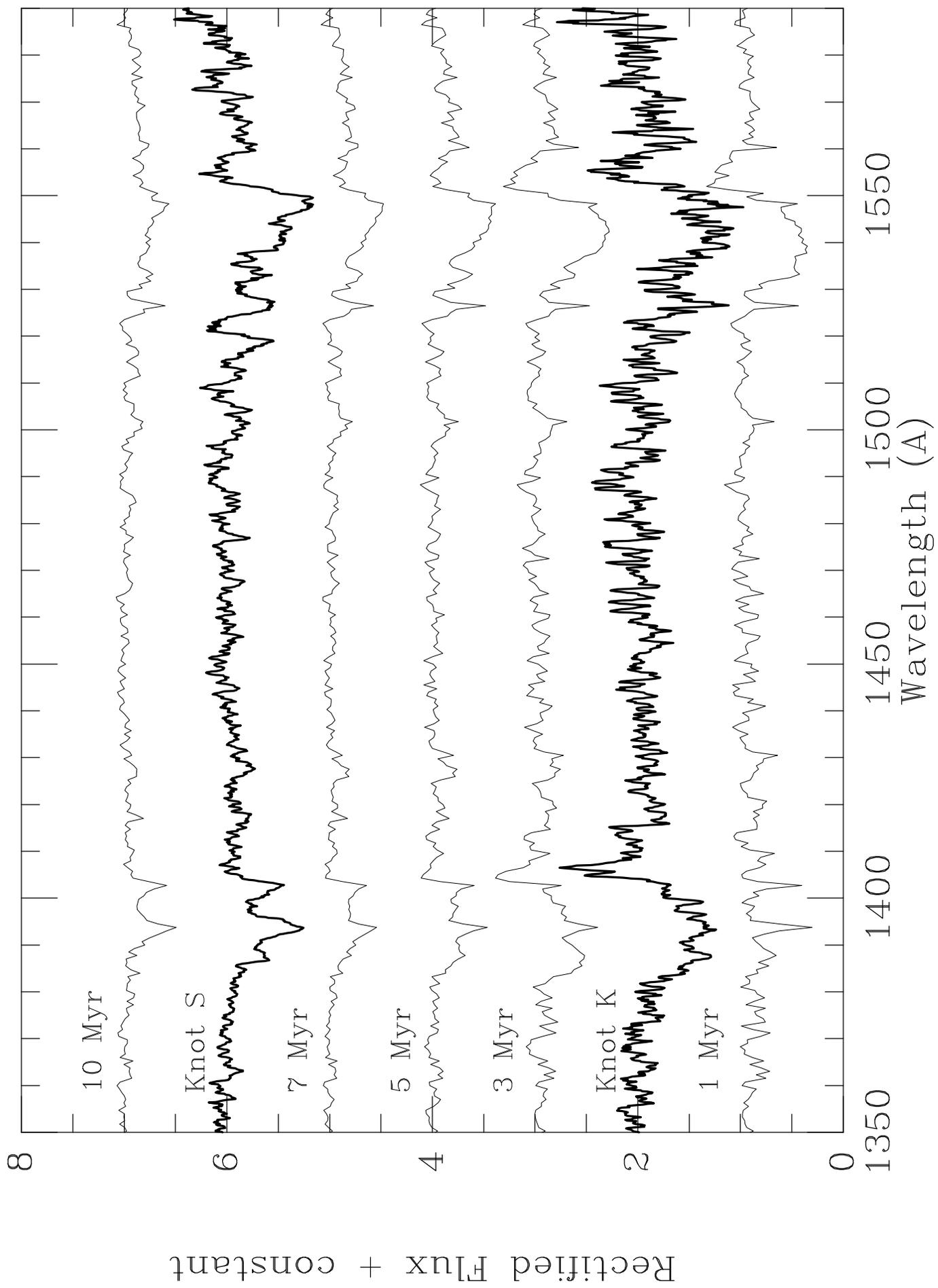
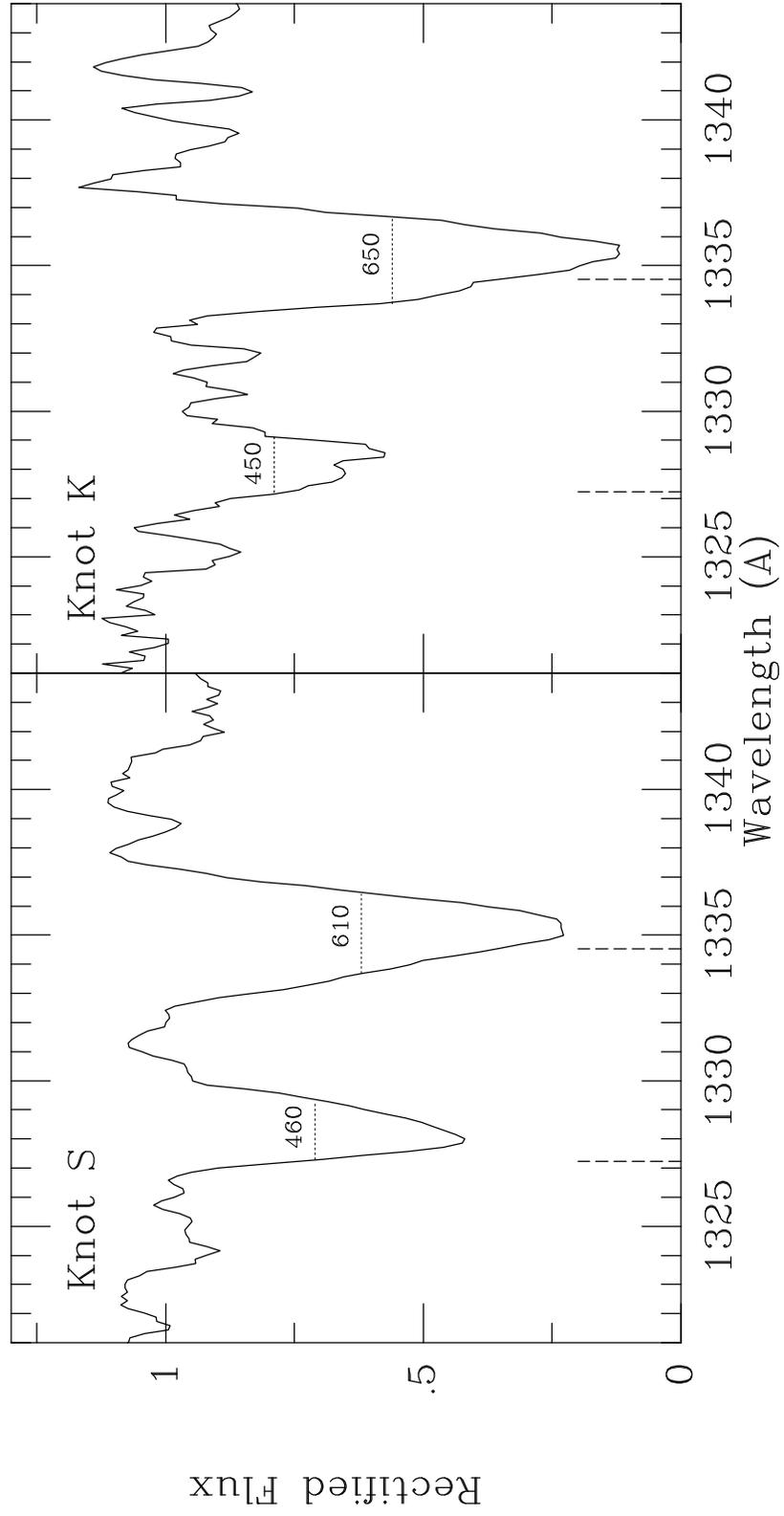
\end{document}